\documentclass[a4paper,UKenglish,cleveref, autoref, thm-restate]{lipics-v2021}

\usepackage{macro}

\title{CFT-Forensics: High-Performance Byzantine Accountability for Crash Fault Tolerant Protocols} 

\author{Weizhao Tang}{Carnegie Mellon University, USA}{wtang2@andrew.cmu.edu}{https://orcid.org/0009-0002-3676-4582}{}

\author{Peiyao Sheng}{University of Illinois Urbana-Champaign, USA}{psheng2@illinois.edu}{https://orcid.org/0000-0002-1896-2852}{}

\author{Ronghao Ni}{Carnegie Mellon University, USA}{ronghaon@andrew.cmu.edu}{}{}

\author{Pronoy Roy}{Carnegie Mellon University, USA}{pronoyroy.11@gmail.com}{}{}

\author{Xuechao Wang}{HKUST(GZ), China}{xuechaowang@hkust-gz.edu.cn}{https://orcid.org/0000-0001-6918-2699}{This work is supported in part by a gift from Stellar Development Foundation and by the Guangzhou-HKUST(GZ) Joint Funding Program (No. 2024A03J0630).}

\author{Giulia Fanti}{Carnegie Mellon University, USA}{gfanti@andrew.cmu.edu}{https://orcid.org/0000-0002-7671-2624}{}

\author{Pramod Viswanath}{Princeton University, USA}{pramodv@princeton.edu}{https://orcid.org/0000-0003-3171-8667}{This work is supported in part by NSF CNS-2325477, ARO W911NF2310147 and C3.AI.}

\funding{
    \emph{Weizhao Tang, Ronghao Ni, Pronoy Roy and Giulia Fanti}: This work was supported in part by the National Science Foundation under grants CNS-2325477, CIF-1705007, and CCF-2338772, and the Air Force Office of Scientific Research under award number FA9550-21-1-0090. 
    We also thank Chainlink Labs, Ripple Labs, and IC3 industry partners for their generous support, 
    as well as Bosch, the Sloan Foundation, Intel, and the CyLab Secure Blockchain Initiative.
}

\authorrunning{W. Tang, P. Sheng, R. Ni, P. Roy, X. Wang, G. Fanti and P. Viswanath} 

\Copyright{Weizhao Tang, Peiyao Sheng, Ronghao Ni, Pronoy Roy, Xuechao Wang, Giulia Fanti and Pramod Viswanath} 



\category{} 

\relatedversion{} 

\acknowledgements{
    We wish to thank Chris Meiklejohn and Heather Miller for their valuable insights and advice on this project. 
    We also thank Sam Stuewe and the MIT Digital Currency Initiative for their feedback and insights regarding integration with OpenCBDC and applications to central bank digital currency. 
}

\nolinenumbers 

\EventLongTitle{}
\EventShortTitle{}
\EventAcronym{}
\EventYear{2024}
\ArticleNo{}

\begin{document}




\maketitle

\begin{abstract}

Crash fault tolerant (CFT) consensus algorithms are commonly used in scenarios where system components are trusted---e.g., enterprise settings and government infrastructure. However, CFT consensus can be broken by even a single corrupt node. A desirable property in the face of such potential Byzantine faults is \emph{accountability}: if a corrupt node breaks protocol and affects consensus safety, it should be possible to identify the culpable components with cryptographic integrity from the node states. Today, the best-known protocol for providing accountability to CFT protocols is called PeerReview; it essentially records a signed transcript of all messages sent during the CFT protocol. Because PeerReview is agnostic to the underlying CFT protocol, it incurs high communication and storage overhead. We propose CFT-Forensics, an accountability framework for CFT protocols. We show that for a special family of \emph{forensics-compliant} CFT protocols (which includes widely-used CFT protocols like Raft and multi-Paxos), CFT-Forensics gives provable accountability guarantees. Under realistic deployment settings, we show theoretically that CFT-Forensics operates at a fraction of the cost of PeerReview. We subsequently instantiate CFT-Forensics for Raft, and implement Raft-Forensics as an extension to the popular nuRaft library. In extensive experiments, we demonstrate that Raft-Forensics adds low overhead to vanilla Raft. With 256 byte messages, Raft-Forensics achieves a peak throughput 87.8\% of vanilla Raft at 46\% higher latency ($+44$ ms). We finally integrate Raft-Forensics into the open-source central bank digital currency OpenCBDC, and show that in wide-area network experiments, Raft-Forensics achieves 97.8\% of the throughput of Raft, with 14.5\% higher latency ($+326$ ms). 

\end{abstract}

\section{Introduction}

In the theory and practice of distributed systems, crash fault tolerance plays a central role \cite{lynch1996distributed}. 
Crash fault tolerant (CFT) protocols allow a system to come to consensus on a log of events even in the presence of nodes that may crash, but otherwise follow protocol \cite{van2015paxos,raft,hyperledger,zookeeper}. 
CFT systems are widely deployed in enterprise systems and support various high-profile services  \cite{hasan2018fault,netflix,burrows2006chubby,zookeeper,garg2013apache}. For example, prevalent systems like etcd \cite{etcd}, CockroachDB \cite{taft2020cockroachdb} and Consul \cite{consul} employ CFT protocols like Raft \cite{raft}.
CFT protocols are also widely-used in security-sensitive critical infrastructure \cite{sakic2017response,hasan2018fault}, including prospective Central Bank Digital Currencies (CBDCs) \cite{lovejoyhamilton,opencbdc-tctl}.

CFT protocols provide theoretical correctness guarantees  under the assumption that at least a certain fraction of nodes follow  protocol, and remaining nodes may suffer from crashes.
However, these assumptions can be broken in practice. 
For instance, an agent could be Byzantine, meaning that it can misbehave arbitrarily, e.g., by delaying or tampering with messages. 
In such cases, consensus can be trivially broken.

One possible solution is to replace the CFT protocol with a \emph{Byzantine fault tolerant} (BFT) protocol, which guarantees consensus under not only crash faults, but also under Byzantine faults \cite{castro1999practical,liu2016xft,cachin2017blockchain,bano2019sok,yin2019hotstuff,gao2022dumbo,gelashvili2022jolteon,guo2022speeding}.  
This is a viable solution, though swapping out consensus protocols may be impractical for organizations that have already built infrastructure around a particular CFT system. 


In this paper, we explore a complementary approach to managing Byzantine faults: \emph{accountability}.
That is,  in the case of Byzantine faults in a CFT protocol, can an auditor with  access to 
locally-stored protocol states 
identify \emph{which} node(s) were responsible for the misbehavior, with cryptographic guarantees? 
In particular, we want to provide this guarantee by making minimal changes to an existing system and protocol, rather than completely replacing the consensus mechanism. 

Accountability for BFT protocols has been studied systematically very recently, both as an intrinsic attribute of existing protocols \cite{sheng2021bft,neu2022availability,neu2023accountable} and as an important feature in the design of new protocols \cite{casper,stewart2020grandpa,civit2019polygraph,sheng2022player}. 
However, there is comparatively little work on CFT protocols that incorporate accountability for Byzantine faults \cite{peerreview,hyperledger}.
An important prior work called PeerReview tackled this problem in the context of general CFT protocols \cite{peerreview}. 
PeerReview works by producing a signed transcript of every message that is sent in the protocol.
Being a general-purpose protocol, it does not always achieve competitive performance with the underlying CFT protocol  (details in \sref{sec:comparison-peerreview}). 
Hence, to our knowledge, existing work on accountability for CFT protocols either: (1) is very general, and thus incurs high performance overhead when applied to specific CFT protocols (i.e., PeerReview \cite{peerreview}), and/or (2) does not include a full implementation-based evaluation to measure the practical effect of accountability \cite{peerreview,hyperledger}. 


\textbf{Our goal in this work is to design a practical accountability framework that incurs low communication and storage overhead by exploiting the structure of the underlying protocol, unlike PeerReview.}
Crucially, despite exploiting protocol structure, we want the framework to be broadly applicable to common CFT protocols 
and backwards-compatible with existing systems. 
To this end, our contributions are threefold:

\begin{figure}
    \centering
    \includegraphics[width=.5\linewidth]{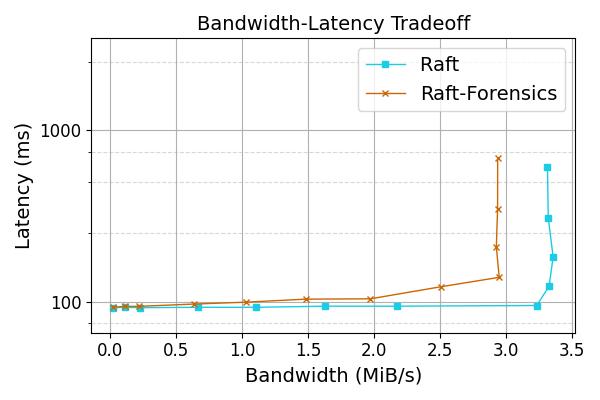}
    \caption{Bandwidth-latency tradeoffs of Raft vs Raft-Forensics over 4 nodes at message size of 256 Bytes.}
    \label{fig:main}
\end{figure}

\begin{itemize}
    \item \textbf{Accountably-Safe Consensus:} We first formally define a subclass of CFT protocols called \familyname protocols, which includes two of the most widely-used CFT protocols in use today: Raft \cite{raft} and Paxos \cite{originalpaxos,heidivs}
    \footnote{{For notational brevity, we use the name `Paxos' to refer to variants of the Paxos algorithm that that are sometimes referred to as multi-Paxos to distinguish from the original single-decree Paxos \cite{originalpaxos,heidivs}.}}. 
    Intuitively, the defining feature of this class is that its protocols cycle between two phases: log replication and leader election, and each phase satisfies some formal properties (defined in Section \ref{sec:model}.  We then  propose \sysname, a lightweight modification to \familyname CFT protocols that \emph{provably} guarantees to expose at least one node that committed Byzantine faults when {consensus is violated. 
    Note that we cannot guarantee to detect more than one Byzantine node, as only one malicious node is needed to break CFT consensus; however, for certain classes of attacks involving multiple Byzantine nodes, we are able to detect multiple misbehaving nodes (\sref{sec:audit-full}).
    } 

    \item \textbf{Theoretical Efficiency Comparison:} We theoretically analyze the communication and computational overhead of \sysname compared to the most relevant prior work in this space, PeerReview. 
    We show that \sysname has (amortized) vanishing storage overhead compared to the baseline protocol in practical scenarios, 
    while PeerReview has overhead that grows linearly with the logs. 
    In addition, during log replication, the communication overhead of \sysname is $58\%$ lower than PeerReview. 
    
    \item \textbf{Empirical Performance Evaluation on Raft:} We implement \raftname, an instantiation of \sysname for the Raft protocol. Our implementation is built on a fork of nuRaft, a popular C++ implementation of Raft. We evaluate its performance compared to Raft, both in benchmark experiments and in a downstream application---specifically, OpenCBDC \cite{lovejoyhamilton}---an open-source central bank digital currency (CBDC) implementation that uses nuRaft.
    {
    In benchmark experiments, we observe in Fig. \ref{fig:main} that \sysname achieves performance close to vanilla Raft 
    (experimental details in \sref{sec:eval}). 
    For instance, in end-to-end experiments, it achieves a maximum throughput that is 87.8\% the maximum throughput of vanilla Raft, at $46\%$ higher confirmation latency (44 ms). 
    In our OpenCBDC experiments over a wide-area network, Raft-Forensics achieves 2.2\% lower throughput at 14.4\% higher latency (326 ms) than vanilla Raft. 
    }

\end{itemize}

\section{Related Work}

\paragraph*{CFT protocols} 
CFT protocols are designed to handle crash faults, where nodes may fail but do not exhibit malicious behavior. 
Paxos~\cite{lamport2019part} is a foundational CFT protocol, with many variants \cite{lampson1996build,originalpaxos,de2000revisiting,lamport2001paxos,boichat2003deconpaxos,abcdpaxos,meling2013tutorial,van2015paxos,heidivs}. 
Raft~\cite{raft} is a CFT protocol that aims to provide a more understandable and easier-to-implement alternative to Paxos \cite{van2015paxos}. 
{Both Raft~\cite{taft2020cockroachdb,roohitavaf2019session,banker2016mongodb} and Paxos~\cite{cassandra2014apache,burrows2006chubby,sivasubramanian2012amazon,corbett2013spanner} are widely-used in practice.}

\paragraph*{Accountability} 
Accountability allows protocols to identify and hold misbehaving participants responsible when security goals are compromised \cite{kusters2010accountability}. 
In the context of fault-tolerant protocols, accountability allows a protocol to identify culpable participants when security assumptions are violated and demonstrate their misconduct. Recent work \cite{sheng2021bft} has examined several widely used BFT protocols and assessed their inherent accountability levels without altering the core protocols.
Since CFT protocols are explicitly designed to handle only crash faults, integrating accountability offers a lightweight enhancement to detect Byzantine actors. %

One prior work \cite{hyperledger} explored the accountability of the Hyperledger Fabric blockchain, which features a pluggable consensus mechanism. This study conducted a case analysis of incorporating accountability into a Hyperledger Fabric system underpinned by a CFT protocol, Apache Kafka \cite{garg2013apache} (called Fabric*). However, this work treats the consensus module as a cluster, offering accountability only at the level of the entire consensus group (not individual nodes within the group). 
In contrast, we aim to identify and attribute Byzantine faults to individual misbehaving consensus replicas participants.  
Fabric* introduces two primary modifications. 
First, parties must sign every message they send. 
Second, it enforces a deterministic block formation algorithm to eliminate ambiguity. 
However, these changes are neither necessary nor sufficient for ensuring accountability in the CFT protocols we study.
In addition, Fabric* does not empirically evaluate their system, whereas we evaluate performance both theoretically and empirically.


PeerReview \cite{peerreview} builds a framework for accountability that applies to general distributed systems. 
Although it accounts for Byzantine faults in CFT protocols as \sysname does, it has substantially higher overhead communications and space requirements than \sysname, which we discuss in \S\ref{sec:perf} in detail. 
PeerReview requires nodes to audit each other, instead of assuming a central auditor as we do (\S\ref{sec:model}).
To address this difference, we disable inter-node auditing in PeerReview, which  
still incurs substantially higher communication and memory overhead than \sysname. 

\section{Setup}

\label{sec:problem}

We study consensus protocols that solve the crash-fault tolerant state machine replication (CFT-SMR) problem over partially synchronous networks. 
Precisely, we consider a setting with $n$ servers (also known as nodes) and arbitrarily many clients. 
For the vanilla CFT-SMR setting, we assume that at most $f$ out of the $n$ nodes can suffer \emph{crash failures}, where they stop working without resuming at an arbitrary and unpredictable moment. 
Each node maintains a state machine \texttt{SM} and an append-only log list \texttt{logs}. 
The goal of CFT-SMR is for the nodes to maintain consistent state machines \texttt{SM} with each other (Definition \ref{def:smr}).
\texttt{SM} maintains a local state $s$ initialized to $s_0 = \perp$ and a deterministic function $\phi$. 
\texttt{logs} are sequential inputs to \texttt{SM} generated from client requests, which results in state transition 
\[
    \mathtt{SM}.s_i = \mathtt{SM}.\phi(\mathtt{SM}.s_{i-1}, \mathtt{logs}[i]), \qquad \forall i \in \mathbb Z_{> 0}.
\]

The network is partially synchronous, meaning that there exists a global stabilization time (GST) and a constant time length $\Delta$, 
such that a message sent at time $t$ is guaranteed to arrive at time $\max\{\mathrm{GST}, t\} + \Delta$. 
GST is unknown to the system designer and is not measurable by any component of the system.

\begin{definition}[CFT(-SMR) Protocol]
    In the setting above, a consensus protocol $\Pc$ is $f$-CFT(-SMR) if $f$ nodes can fail by crash, and the following three properties are satisfied. 
    \begin{enumerate}
        \item \textbf{Safety: If $E$ is the $i$-th entry of a {correct} node's log, then no other correct node has $E' \neq E$ at index $i$. }
        
        \item Liveness: If a {correct} client submits a request $r$, then eventually all non-faulty nodes will (1) have a log entry $E$ at index $i$ handling $r$ (2) there exists a log entry at all previous positions $j < i$.
        
        \item Validity: Each entry in the log of a correct replica can be uniquely mapped to a command proposed by a client request.
    \end{enumerate}
    \label{def:smr}
\end{definition}

In the remainder of the paper, we study $f$-CFT protocols with $f = \floor{(n-1) / 2}$ and focus on the \textbf{boldfaced} safety property. 
These protocols tolerate $f$ crash failures, but are typically vulnerable under even one \emph{Byzantine failure}, where a node arbitrarily deviates from the stipulated protocol (\S\ref{subsec:attacks}). 

We formalize our threat assumptions below. 

\subsection{Threat Model}

In addition to the $f$ nodes with crash failures, we further assume the existence of $b \ge 1$ nodes that execute Byzantine faults. 
We assume $b \le n-2$ to avoid a trivial problem with at most one honest node.
The Byzantine nodes are capable of accessing states of honest nodes and collaboratively determining whether, when, and what to send to every honest node. 
However, they cannot influence the honest nodes or the communication between them.  

\paragraph*{Auditor}
To identify the adversary, we introduce an \emph{auditor} in addition to the clients and servers in the SMR model.
The auditor may query the full states of any node 
(details in \S\ref{sec:state}). 
If an auditor requests information, honest nodes always provide their information to the auditor; a Byzantine node can respond arbitrarily. 
The auditor determines the safety of the system by checking data legitimacy and consistency among the nodes, as a function of the received state information.  
However, auditors are unable to directly influence the system. 
Our main goal is to define modifications to the consensus protocol and an auditing algorithm that jointly enable an auditor to uncover the identity of the adversarial node if the state machine safety property is violated or a legitimate client receipt conflicts with the state machine logs.

\subsection{The Accountability Problem}

If even a single node is Byzantine, CFT protocols are vulnerable to safety violations (examples in Section \ref{subsec:attacks}). 
As a result, we want to identify the party responsible for a safety violation using an auditing algorithm. If such an algorithm exists, we say the protocol has \emph{accountability}. 

\begin{definition}[Accountability]
Let $\Pc$ denote a $f$-CFT-SMR consensus protocol. 
$\Pc$ has \emph{accountability} if there exists a polynomial-time auditing algorithm $\Ac$ s.t. 
\begin{enumerate}
    \item $\Ac$ takes the states of $\Pc$ as input. 
    \item If safety (Def. \ref{def:smr}) is violated, $\Ac$ outputs a non-empty set of nodes and irrefutable proof {that each member of the set violated protocol.}
    Otherwise, $\Ac$ outputs $\perp$. 
\end{enumerate}
\label{def:acc}
\end{definition}



\section{\FamilyName Protocols: A Family of CFT Protocols}
\label{sec:model}

Modifying an arbitrary CFT-SMR protocol under a general workflow without context can be challenging. 
To address this, we define a family of CFT-SMR consensus protocols named \emph{\familyname}, 
which are provably modifiable for accountability under our general framework \sysname (Def. \ref{def:acc} and Theorem \ref{thm:cft-acc}). 
At a high level, a \familyname protocol is leader-based (Property \ref{prop:leader}). 
It can be described by a set of \emph{procedures}, which is partitioned into \emph{log replication} and \emph{leader election}\footnote{We use the terminology of Raft for clarity.} with each satisfying necessary properties. 
Both Raft \cite{raft} and \Multipaxos \cite{van2015paxos},  two dominant CFT protocols in practice \cite{heidivs}, are \familyname protocols.

\paragraph*{Setup}


We start with a $f$-CFT-SMR protocol. 
In the protocol, each entry in the log has two possible states: \textit{committed} and \textit{uncommitted}. If an entry is committed, the content in the entry will not be changed in the future and can be applied to the state machine. If a prefix in the log is committed, then all entries in the prefix is considered committed. The largest index of committed entries is called \emph{the last commit index}, denoted as \lci.

Let there be global notion of time $T = [0, \infty)$, which is unknown to any of the nodes. 
For simplicity, let $x[i]$ denote the $i$-th log entry in node $x$'s log list. 
For a given entry $E$ with index $i$, we say a node $y$ \emph{owns} $E$ if $y[i] = E$. 
Furthermore, we let $E_{i:j}$ denote a sequence of consecutive entries $\{E_k|k \in [i, j],~ E_k\dindex = k\}$.
Throughout the paper, we use \method{colored~monospace} text to denote protocols and methods that appear in pseudo-code.  


\paragraph*{Leader-Based}

A \familyname protocol must  satisfy the \emph{leader-based property} (Property \ref{prop:leader}). 

\begin{property}[Leader-Based]
\label{prop:leader}
    At any time $t \in T$, each node $x \in S$ identifies a leader $L_x(t) \in \bar S \triangleq S \cup \{\perp\}$.
    For each $x$, there exists an interval partition of $T = \bigcup_{i=1}^\infty [t_{i-1}, t_i)$ and a sequence of nodes $\{\ell_i \in \bar S\}_{i=1}^\infty$ where for all $i \in \mathbb Z_{>0}$, 
    \[
        t_{i-1} < t_i, \quad \ell_i \neq \ell_{i+1}; \quad L_x(t) = \ell_i, \forall t \in (t_{i-1}, t_i). 
    \]
    
    If $L_x(t) = x$ for all $t \in [\underline{t}, \bar t)$, $x$ is called a \emph{leader} during the leadership $[\underline{t}, \overline{t})$. 
    Otherwise, if $L_x(t) = \ell \notin \{x, \perp\}$, $x$ is called a \emph{follower} identifying $\ell$. 
    Only a leader can propose a log entry. 
    At time $\underline{t}$ when $\ell$ starts being a leader, 
    it assigns a unique term to itself which is fixed until it stops being a leader at $\bar t$.
    Hence, the term can be regarded as an attribute of a leadership during $[\underline{t}, \overline{t})$.
    For node $x$ to identify $\ell$, $x$ must receive a message from $\ell$ that includes $\ell$'s term. 
    $x$ sets its term equal to $\ell$'s term as soon as it starts identifying $\ell$. 
    
    We say there exists a \emph{global leader} $\ell \in S$ of term $\tau$, if there exists a majority subset $M \subseteq S$, such that $\ell \in P$ and for all $x \in M$, $x\dterm = \tau$. 
    Since $M$ is the majority, $\ell$ must be unique at every time, so global leaderships do not overlap in time. 
    We require that the term of a later global leadership must be strictly greater than that of an earlier one. 
\end{property}

The full protocol consists of \emph{procedures} that are partitioned into the following subprotocols. 

\begin{itemize}
    \item \emph{Log Replication} is the subprotocol that collects all procedures only executed when the host node $x$ identifies a new leader, i.e., $L_x(t) \neq \perp$.
    \item \emph{Leader Election} is the subprotocol collecting all the remaining procedures. 
\end{itemize}

\paragraph*{Log Replication}

{
On the top level, log replication (Alg. \ref{alg:general-lr}) has a main procedure \method{HandleClientRequest} that is triggered when a leader receives a client request.
If a node is not running \method{HandleClientRequest} or involved in an RPC call within the procedure, it cannot
create a new log entry or edit its logs and \lci.   
It has three steps -- log entry creation, replication and commitment. 
}

\begin{figure}
    \centering
    \includegraphics[width=\linewidth]{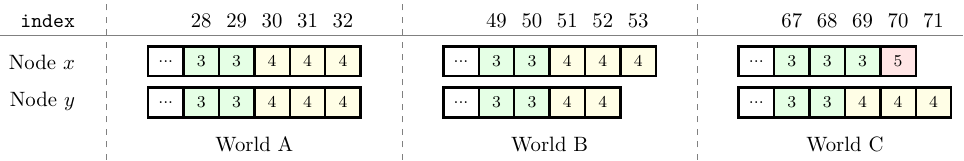}
    \caption{Examples of node freshness. Each box represents an entry containing the entry's \texttt{term}. In all worlds, $x$'s log list is fresher than $y$'s. In World A,  $x$ is as fresh as $y$. In only Worlds B and C, $x$ is \emph{strictly} fresher than $y$. }
    \label{fig:fresh}
\end{figure}

\emph{Creation.} When leader $\ell$ receives a client request, $\ell$ creates a corresponding log entry $E$ and appends it to the log list. $E$ has 3 attributes -- 
(1) \code{term}, $\ell$'s term; 
(2) \code{index}, its index on the log list; and 
(3) \code{payload}, which handles the request. 
We define the \emph{freshness} of a log entry, a log list and a node in Def. \ref{def:fresh}, and provide an example in Fig. \ref{fig:fresh}. 

\begin{definition}[Freshness]
    A log entry $E$'s freshness is denoted by the tuple $(E\dterm, E\dindex)$. 
    $E$ is \emph{as fresh as} entry $F$ if their freshness tuples are identical.    $E$ is \emph{fresher} than $F$ if 
    $ E\dterm > F\dterm$ or $E\dterm = F\dterm \wedge E\dindex \ge F\dindex$.
    $E$ is \emph{strictly fresher} than $F$ is $E$ is fresher than $F$, and $E$ is not as fresh as $F$. 
    In contrast, $E$ is \emph{staler} than $F$ if $E$ is not strictly fresher than $F$. 
    The freshness of a node or its log list is equivalent to that of the log list's last entry. 
    \label{def:fresh}
\end{definition}

\emph{Replication.} The procedure of replication can be described by $\ell$ calling a single RPC $\method{AppendEntries}$ for each remaining node. 
Its eventual outcome is a \code{AppendEntriesResp} message from each callee, which includes a predicate \texttt{accept} that indicates whether the replication is successful. 
In addition, the RPC must satisfy the replication property:
\begin{property}[Replication]
    If a follower $x$ replicates $E_{i:j}$ from leader $\ell$, 
    $x$'s term must equal $E_j$'s term, and it must own $E_{1:i-1}$. 
    Formally, $x\dterm = \ell\dterm = E_j\dterm$ and for all index $k \le j$, $x[k] = \ell[k]$.
    \label{prop:repl}
\end{property}

\emph{Commitment.} Once $\ell$ receives  
\code{AppendEntriesResp} messages {from $(n-f-1)$ followers} with \texttt{accept=True}, $\ell$ commits $E_j$. 
Then, $\ell$ sends a message \code{InformCommitMsg} including (hash of) $E_j$ to each remaining node $x$, who also commits $E_j$ if it owns $E_j$. 

\begin{algorithm}[!htb]
\caption{Log replication of the \familyname family.  
In persistent storage, a node maintains \texttt{term}, \texttt{logList}, \lci, \fmod{\ccert and \lcert}, 
where \lci is the last commit index. 
The \fmod{red lines and variables} are added in  CFT-forensics (\S\ref{sec:cftforensics}). 
}
\label{alg:general-lr}

\Pn{\FLR{host node $z$}}{
    \textbf{As leader:}\;
    \PDn{\method{HandleClientRequest}($r$)} {
        
        $i \leftarrow \texttt{logList}\dlen + 1$\tcp*{Creation}
        $E \leftarrow $ \code{LogEntry}\texttt{(term=term, index=$i$, payload=\method{Payload}($r$))}\;
        \color{modclr}
        $E\dptr \leftarrow \method{Hash}(i\|E\dpay\|\texttt{logList}[i-1]\dptr)$\;
        $E\dsig \leftarrow \sigma_z(E\dptr)$\;
        $E\texttt{.\lcert} \leftarrow \lcert$\;
        \color{black}
        
        Append $E$ to \texttt{logList}\;

        \texttt{replicators}, \fmod{\texttt{sigs}} $\leftarrow \{z\}, \fmod{\{\sigma_z(E\dptr)\}}$\tcp*{Replication}
        \For(){async $x \in S - \{z\}$} {
            $\msg : \mathtt{AppendEntriesResp} \leftarrow$  async \method{AppendEntries}($[E]$, ...)\;
            \If{\texttt{msg.accept}}{
                \color{modclr}
                \If{\texttt{verifySig}(\texttt{msg.signature})} {
                    \textbf{fail exit}\;
                }
                $\texttt{sigs} \leftarrow \texttt{sigs} \cup \{\msg\texttt{.signature}\}$\;
                \color{black}
                $\texttt{replicators} \leftarrow \texttt{replicators} \cup \{x\}$\;
            }
        }
        Wait until $|\texttt{replicators}| \ge n-f-1$\;

        \If(\tcp*[f]{Commitment}){$E$ not yet committed} {
            $\method{Commit}(E)$\;
            
            \color{modclr}
            $\ccert \leftarrow i \| h_i \| \texttt{replicators} \| \mathtt{sigs}$\; 
            \color{black}
        
            \For{$x \in S - \{z\}$} {
                Send $\code{InformCommitMsg}(E, \fmod{\ccert})$ to $x$\; 
            }
        }
    }
}
\end{algorithm}

\paragraph*{Leader Election}

By definition, leader election is the set of procedures that do not belong to Log Replication, where \method{CandidateMain} is the main procedure. 
Without running \method{CandidateMain} or being involved in an RPC within it, a node cannot identify any leader. 
Only a candidate within \method{CandidateMain} can edit its logs and \lci. 
\method{CandidateMain} consists of three steps: term switching, candidate qualification, and leadership claim.

\emph{Term Switching.} 
At the beginning, the caller $\ell$, also called a \emph{candidate}, updates the term to a greater term, which is exactly the term of $\ell$'s leadership as the outcome of \method{CandidateMain}. 

\emph{Candidate Qualification.}
This phase is represented by a procedure \method{Qualification}, 
which can be completed or \emph{interrupted}.
If it is interrupted, \method{CandidateMain} is also interrupted. 
Otherwise, it satisfies the election property (Property \ref{prop:election}) by necessary communications and state modifications. 

\begin{property}[Election]
    If $\ell$ completes \method{Qualification} at term $\tau$, there must exist a set of nodes $X$ where $|X| \ge n-f$ and $\ell \in X$, such that 
\begin{enumerate}
    \item (Validity) After \method{Qualification}, $\ell[j]\dterm \ge \ell[i]\dterm$ for all $j > i$. 
    
    \item (Selection) For every log entry $E=\ell[i]$ \textbf{after} \method{Qualification}, there exists $x \in X$ such that $x[i]\dpay = E\dpay$.
    
    \item (Freshness) Let $x^i$ denote an arbitrary node satisfying Selection at index $i$.  
    Before \method{Qualification}, let node $y$ be freshest among $X$.
    After \method{Qualification}, $\ell$'s log list is no shorter than $y$'s and for all $i \le $ length of $y$'s log list, $x^i[i]\dterm \ge y[i]\dterm$. 
\end{enumerate} 
    \label{prop:election}
\end{property}

\emph{Leadership Claim.}
After \method{Qualification}, $\ell$ identifies itself as the leader. 
Then, it sends a \code{LeadershipClaim} message including its term $\tau$ to each other node. 
A recipient $x$ identifies $\ell$ as the leader and sets its own term to $\tau$ if $\tau$ is greater than $x$'s own term; 
otherwise, $x$ ignores the message.

\begin{algorithm}[!htb]
\caption{Leader election of the accountable family. 
\fmod{The red lines and phrases are specific to our  (unoptimized) CFT-forensics (\S\ref{sec:cftforensics}).}
}
\label{alg:elect-general}

\Pn{\FLE{host node $z$}}{
    \PDn{\method{CandidateMain}()} {
        $\texttt{term} \leftarrow $ a new, higher term than \texttt{term} \tcp*{Term switching}

        \method{Qualification}(\texttt{term}, ...) \tcp*{Candidate qualification}

        \color{modclr}
        $r \leftarrow z \| \texttt{term} \| z[\mend]\dterm \| \mend \| z[\mend]\dptr$\;
        $\mathtt{votes} \leftarrow \{ z: \sigma_z(\method{Hash}(r))\}$\;
        \For(){async $x \in S - \{z\}$} {
            \msg $\leftarrow$ Call RPC $\method{RequestVote}(x, \texttt{term})$\;
            \If{\texttt{verifySig}(\texttt{msg.signature})} {
                $\mathtt{votes}[x] \leftarrow  \texttt{msg.signature} $\;
            }
        }
        Wait for $\texttt{votes.size} \ge n-f$\;
        
        $\lcert \leftarrow r \| \texttt{votes.keys} \| \texttt{votes.values} $\;
        \For{$E \in \texttt{logList}$} {
            \If{$E\dterm = \texttt{term}$} {
                $E\texttt{.LC} \leftarrow \lcert$\;
            }
        }
        \color{black}
        
        $L_z \leftarrow z$ \tcp*{Leadership Claim}
        \For(){$x \in S - \{z\}$} {
            Send $\texttt{LeadershipClaim}(\texttt{term}, \fmod{\lcert})$ to $x$\;
        }
    }
    \Rn{\method{HandleClaimLeadershipMsg}($\ell$, \msg)} {
        \If{$\mathtt{msg}\dterm \le \texttt{term}$} {
            \textbf{fail exit}\;
        }
        $L_z, \texttt{term} \leftarrow \perp, \texttt{msg}\dterm$\;
        \color{modclr}
        \If{\KwNot\texttt{validate(msg.\lcert)}} {
            \textbf{fail exit}\;
        }
        \color{black}

        $L_z \leftarrow \ell$\;
        ...\;
    }
}
\end{algorithm}


\subsection{Summary}

\begin{definition}[\FamilyName Protocols]
A \familyname protocol is a leader-based (Property \ref{prop:leader}) $f$-CFT-SMR protocol (Def. \ref{def:smr}). 
The protocol can be partitioned into two subprotocols -- log replication (Alg. \ref{alg:general-lr}) and leader election (Alg. \ref{alg:elect-general}), such that
\begin{itemize}
    \item Log replication is a set of procedures that can only be executed when a node identifies a leader. 
    If a node identifies itself, it handles client requests with \method{HandleClientRequest}, where the \method{AppendEntries} RPC must have the replication property (Property \ref{prop:repl}).
    
    \item Leader election is the set of all the remaining procedures, including \method{CandidateMain}.
    \method{CandidateMain} uniquely allows a node to start identifying a leader. 
    In \method{CandidateMain}, the \method{Qualification} procedure must satisfy the election property (Property \ref{prop:election}). 
\end{itemize}

In addition, the log list and \lci \textbf{must not be modified by any procedure that is not mentioned above or explicitly written in the pseudocode. }
\label{def:family}
\end{definition}

\begin{proposition}[Instances of \FamilyName Protocols]
    Both Raft \cite{raft} and \Multipaxos \cite{originalpaxos,heidivs} are \familyname.
\end{proposition}

\paragraph*{Proof (Raft)}
\label{sec:raft}
Raft is originally designed in a very similar philosophy to the \familyname family.
It is a leader-based (Property \ref{prop:leader}) SMR solution by design. 
The Raft consensus algorithm has two components: log replication and leader election. 
In detail, Alg. \ref{alg:raft-impl} in \S\ref{sec:raft-paxos-code} implements the core procedures including \method{AppendEntries} and \method{Qualification}. 

\method{AppendEntries}: 
After a follower receives a list of consecutive log entries (or a single entry), it replicates them if it has the predecessor of the head of the list. 
Otherwise, it triggers \method{AppendEntries} recursively to synchronize all uncommitted entries, which guarantees the no-gap property (Property \ref{prop:repl}).

\method{Qualification}: 
A candidate $\ell$ in Raft asks voters for votes, and a voter only votes if $\ell$'s log list is fresher than its own. 
This ensures $\ell$ is fresher than $n-f$ nodes without changing its logs, so \method{Qualification} RPC satisfies the election property (Property \ref{prop:election}). 

To summarize, all the RPCs have the required properties, so Raft is \familyname. 

\paragraph*{Proof (\Multipaxos)}
\label{sec:paxos}
\Multipaxos is an optimized protocol based on a simple array of basic Paxos. 
Its description varies from paper to paper, so we adopt the version in \cite{heidivs} which enables a clear comparison to Raft.
Both the original \Multipaxos\cite{originalpaxos} and \cite{heidivs}'s variation are leader-based (Property \ref{prop:leader}). 
In \Multipaxos, the log replication procedures are identical to those in Raft. 
Thus, we focus on the leader election subprotocol (Alg. \ref{alg:paxos-impl} in \S\ref{sec:raft-paxos-code}). 

Unlike Raft, a \Multipaxos voter $x$ always votes for a candidate $\ell$ with a higher term in \method{Qualification}. 
The vote comes with all $x$'s entries at $\ell$'s uncommitted indices of $\ell$. 
With $n-f-1$ such votes, at each uncommitted index, $\ell$ selects the freshest entry it has ever seen.
Hence, \method{Qualification} in \Multipaxos also satisfies the election property (Property \ref{prop:election}). 

To summarize, \Multipaxos (as described in \cite{heidivs}) is also \familyname. 

\section{CFT-Forensics}

\label{sec:cftforensics}

Although CFT protocols guarantee safety against crash faults,
they are not safety-resilient against even a single Byzantine fault. 
We first illustrate typical safety attacks.
Next, we present \sysname to endow \familyname protocols with accountability. 

\subsection{Example Attacks}
\label{subsec:attacks}

Recall that in \S\ref{sec:problem}, we assumed that $b \in [1, n-2]$ nodes may behave adversarially. 
In two examples, we assume $n = 2f+1$ is odd for simplicity. 
We show the capabilities of a single attacker Mallory, and the remaining $2f$ nodes are evenly partitioned into $X$ and $Y$. 


\begin{example}[Proposer's Attack, or Split-Brains]
\label{eg:log-manip}
    Let Mallory be a corrupt leader. 
    At the same index, Mallory replicates log entries $E$ and $E'\neq E$ to $X$ and $Y$, respectively. 
    At each side, she commits the corresponding entry with a quorum of $f+1$ nodes. 
    As a result, the honest nodes in $X$ and $Y$ have different committed log entries at the same index. 
\end{example}

\begin{example}[Voter's Attack]
\label{eg:obs-voting}
    Let Mallory be a corrupt voter who has committed entry $E$ with $Y$. 
    Nodes in $X$, however, do not own $E$. 
    In an election, suppose Carol $\in X$ who earns all $f$ votes from $X$. 
    When Carol requests vote from Mallory, Mallory votes under simulation of a clone of Carol.
    After being elected, Carol commits $E'\neq E$ at the same index, which conflicts with any honest node in $Y$.
\end{example}

Surprisingly, these two examples almost exhaustively enumerate the types of safety attacks against \familyname protocols {(Theorem \ref{thm:cft-acc})}. 
This is why \familyname protocols can achieve accountability with much simpler modifications than PeerReview. 


\subsection{\sysname Design Overview}

We present \sysname, a framework that enables accountability for  \familyname protocols. 
Here, we present a basic variant of \sysname, which adds large overhead compared to vanilla CFT protocols; we provide and analyze an optimized variant in \S\ref{sec:stateopt}.  
We use the convention that for a \familyname protocol $\Pc$, $\Pc$-Forensics denotes the protocol $\Pc$ augmented with \sysname (e.g., we implement Raft-Forensics in Section \ref{sec:eval}). 

At a high level, \sysname adds two central data structures to a \familyname protocol: commitment certificates (\ccerts) and leader certificates (\lcerts). 
A \ccert irrefutably proves that a quorum of nodes have replicated an entry, 
and an \lcert proves {\emph{which}} quorum of nodes agreed to elect a leader. 
\sysname requires each log entry to be signed by its proposer, which provides accountability for a split-brains attack (Example \ref{eg:log-manip}). 
It also requires that each voter signs its vote, for which the voter is forced to take responsibility since the vote exists in a \ccert or an \lcert,
 providing accountability for the voter's attack (Example \ref{eg:obs-voting}). 

\paragraph*{Additional Assumptions: Public Key Infrastructure}
We assume access to a Public Key Infrastructure (PKI).
Each node $x$ has a pair of private and public keys, 
where the public key is well known, that is, known by all parties in the system, including other nodes and auditors. 
Node $x$ can use its private key to create an unforgeable signature on (the hash of) an arbitrary message $m$, denoted by $\sigma_x(\method{Hash}(m))$, and the signature can be verified with $x$'s public key. 
A collision-resistant cryptographic hash function \FHash is  known to all parties. 
Both signing a message and verifying a signature can be executed in time that is polynomial in message size.


\subsubsection{Added States}
\label{sec:state}
We first explain the new state that is maintained in \sysname.
\sysname introduces four new categories of states: hash pointer, proposer stamp, leader certificate (\lcert) and commitment certificate (\ccert). 

\begin{table}[!htb]
    \centering\small
    \begin{tabular}{cccccc}
    \hline
         \multirow{2}{*}{Commitment certificate \ccert} & & \code{index}&   \code{pointer}& \code{voters}&\code{signatures}\\ 
         && $i$& $h_i$& $V$ & $\{\sigma_x(h_i)\}_{x \in V}$  \\
    \hline
        \multirow{2}{*}{Leader certificate \lcert} &&&\code{req}& \code{voters}&\code{signatures}\\ 
         &&& $r$ & $V$&$\{\sigma_x(\method{Hash}(r))\}_{x \in V}$\\
    \hline
        \multirow{2}{*}{Vote Request} & \code{id}& \code{term} & \code{eterm} & \code{end} & \code{pointer}\\ 
        & $\ell$ & $\ell\dterm$ & $\tau$ & $i$ & $h$ \\
    \hline
        
    \end{tabular}
    \caption{Attributes of a \ccert, a \lcert and a vote request.}
    \label{tab:attr}
\end{table}

\begin{enumerate}[leftmargin=*]
    \item \textbf{Hash Pointer.} 
    The hash pointer of log entry $E_i$ is denoted by $E_i\dptr$, where $E_0\dptr = \perp$. 
    It is a lightweight proof that the host node owns the entire log list from $E_1$ to $E_i$. 
    The other hash pointers can be derived by 
    \begin{align}
        E_{i}\dptr = \FHash(i \| E_{i}\dpay \| E_{i-1}\dptr), \forall i \in \mathbb{Z}_{>0}.
        \label{eqn:hashptr}
    \end{align}
    
    \item \textbf{Proposer Stamp.}
    The (proposer) stamp of log entry $E$ is a digital signature by its proposer $\ell$ on the hash pointer of $E$. We denote it by $E\dsig = \sigma_\ell(E\dptr)$.
    Should a pair of stamps of $E$ and $E' \neq E$ exist where $E$ is neither an ancestor or descendent of $E'$ and $E\dterm = E'\dterm$, $\ell$ must have launched a split-brains attack. 
    
    \item \textbf{Leader Certificate (\lcert) of Proposer.}
    The \lcert of log entry $E$, denoted by $E\texttt{.LC}$, is the \lcert created by $E$'s proposer $\ell$ at term $E\dterm$. 
    It collects a quorum of signatures from a set of nodes $V$ on $\ell$'s vote request, where a request includes ID $\ell$, term $\ell\dterm$, plus (term, index, hash pointer) of $\ell$'s last entry $(\tau, i, h)$. 
    Formally, $r = \ell \| \ell\dterm \| \tau \| i \| h$ and $\lcert = r \| V \| \{\sigma_x(\method{Hash}(r))\}_{x \in V}$, as shown in Table \ref{tab:attr}. 
    
\end{enumerate}

\begin{figure}[!htb]
    \centering
    \includegraphics[width=\linewidth]{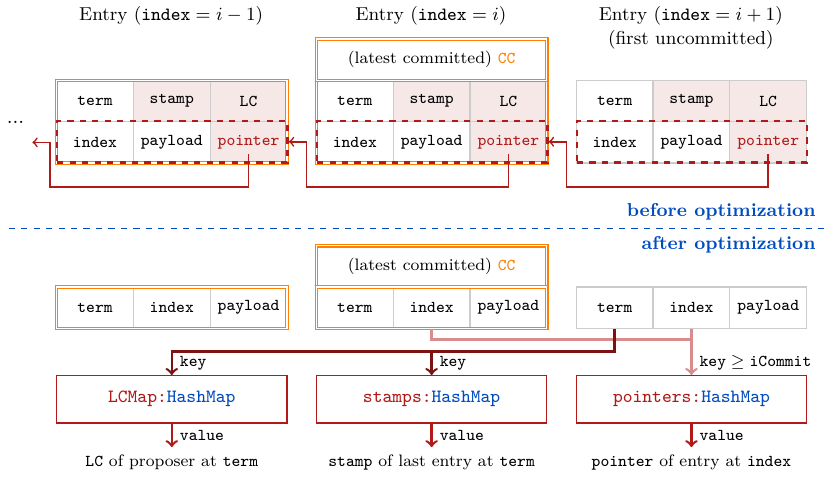}
    \caption{Log entry attributes with and without \sysname; {committed blocks are shown with a double gold outline}. Our basic (unoptimized) \sysname (top) adds a hash pointer, a proposer stamp, and a leader certificate \texttt{LC}, all shown in \fmod{red}. {We also store a \ccert only for the latest committed block. Our optimized \sysname (\S\ref{sec:stateopt}) reduces storage costs by storing three hash maps: (1) one containing pointers only for the last committed block and later uncommitted blocks, (2) one storing a single leader certificate \lcert for every term, and (3) one storing a proposer stamp only for the latest proposed block in the current term.}}
    \label{fig:entry-field}
\end{figure}

In summary, in our basic (un-optimized)  \sysname, a log entry has six attributes (Fig. \ref{fig:entry-field}) -- \code{term}, \code{index}, \code{payload}, \code{pointer}, \code{stamp} and \code{LC}. 
In addition, \sysname requires each node to maintain two independent states -- 4) the current leader's \lcert and 5) the latest \ccert. 

\begin{enumerate}[leftmargin=*]
\setcounter{enumi}{3}
    \item \textbf{Leader Certificate of Current Leader.} Each node additionally maintains the \lcert of the current leader it identifies. 
    This \lcert is not covered above because the current leader may have not proposed any log entry yet. 
    
    \item \textbf{Commitment Certificate (\ccert)}
    Each node only maintains one freshest \ccert.  
    Like \lcerts, a \ccert is a collection of a quorum of signatures on the same log entry. 
    Formally, for a log entry at index $i$ that is replicated to a set of nodes $V$ where $|V| \ge n-f$, we construct a \ccert following the structure in Table \ref{tab:attr}. 
    We denote $\ccert = i \| h_i \| V \| \{\sigma_x(h_i)\}_{x \in V}$. 
\end{enumerate}

\subsubsection{Modified Procedures}

\paragraph*{Log Replication}
We mark our changes in \textcolor{modclr}{red} in Alg. \ref{alg:general-lr}. 
Upon creation of a log entry $E$ at index $i$, the leader $\ell$ correctly attaches the three new states (\code{pointer}, \code{stamp} and \code{LC}). 
Then it replicates the ``enhanced'' entry to followers via the \method{AppendEntries} RPC. 
Upon receipt, each follower $x$ validates the new states, and eventually puts the entries at their correct indices. 
As a result of a successful replication, $x$ sends a \code{AppendEntriesResp} message, which not only includes the predicate \texttt{accept}, but also $x$'s signature on the last entry $E$'s hash pointer. 

With $(n-f-1)$ \code{AppendEntriesResp} messages, the leader updates its \ccert by assembling the $n-f$ signatures it has obtained (including its own). 
To notify followers to commit $E$, the leader sends a \code{InformCommit} message which includes \ccert in addition to $E$. 
Upon receipt, a follower commits $E$ if it owns $E$ and the \ccert passes a follower's verification. 

\paragraph*{Leader Election}
In the \method{Qualification} procedure which satisfies the election property (Property \ref{prop:election}), 
if a candidate $\ell$'s logs are changed during \method{Qualification}, 
we let $\ell$ reconstruct every uncommitted entry with the same payload, as if $\ell$ plans to repropose them. 
In detail, $\ell$ a) sets their terms equal to its current term, b) re-derives their hash pointers, c) creates its own stamp for each of them, and d) sets their proposer LCs to its own LC. 
As a result, the hash pointers will still be correct, and no entry will be overwritten if it has been committed by any node. 

Assume that the candidate $\ell$ passes the \method{Qualification} procedure in a vanilla \familyname protocol. 
Instead of directly declaring leadership in vanilla, $\ell$ broadcasts another vote request $r$ based on its current last log entry by calling \code{RequestVote} RPC. 
Since $\ell$ is already qualified, the request deserves at least $n-f$ votes by election property. 
Each vote from $x$ contains a signature $\sigma_x(\FHash{r})$, 
proving $x$'s awareness that $\ell$ is fresher than itself. 
After collecting $n-f$ votes, $\ell$ assembles a leadership certificate (\lcert) and claims leadership by broadcasting it.
Then, each recipient will verify the \lcert, store it, and identify $\ell$ as the leader. 

In general, we add an additional round of communication to leader election, 
where the candidate provides information of its last log entry and the voters send signatures. 
In \emph{passive} leader elections like \multipaxos, any arbitrary node can be elected under deterministic logic (e.g., under round robin or maximum ID).
The new leader must ensure freshness by updating its log entries based on those it receives from the other nodes. 
As a result, the last log entry is only available after a round of communication, so a second round of signatures is needed. 
However, it is not needed in \emph{active} elections like Raft, 
where a node actively seeks leadership candidacy. 
If each node never modifies its logs during election, then their last entry does not change, and they can collect  signatures in just one round of communication. 

\subsection{Accountability Guarantee}

\begin{theorem}
    If a CFT protocol $\Pc$ is \familyname, then $\Pc$-Forensics achieves accountability (Def. \ref{def:acc}). 
    \label{thm:cft-acc}
\end{theorem}





\textbf{Proof Sketch.} (Full proof in Appendix \ref{sec:proof-safety})
We first establish a map from each term to the \lcert of that term's leader. 
If a term is associated with \lcerts, we can accuse all voters of both leaders for voting twice at the same term. 
If this map exists, a term is uniquely used by a leader. 
Since safety (Def. \ref{def:smr}) does not hold, we find the first pair of entries from the logs of two honest nodes that conflict.

If they are of the same term, we discover a \emph{split-brains} attack and we can accuse the leader by its stamps on the conflicting entries or their successors. 

If they are of different terms, we discover a voter's attack, which has two possibilities -- 
1) at least one voter voted for a leader not fresher than itself; and 
2) at least one voter replicated and signed an entry at a term less than its term. 
In this final case, we can accuse all the voters who have signatures in a pair of conflicting \ccert and \lcert. \qeda

\section{Performance Comparison with PeerReview}
\label{sec:perf}

In this section, we provide a head-to-head comparison of the theoretical overhead costs of \sysname compared to PeerReview, for the special cases of \raftname and \paxosname. 
We begin by explaining some practical optimizations that reduce the redundancy of \sysname without affecting accountability, then explain the cost comparison calculations. 

\subsection{\sysname State Optimization}
\label{sec:stateopt}
The added states in basic \sysname  incur linear overhead in the number of log entries. 
We next show how to store the new states in independent,  more efficient data structures. 

\textbf{Hash Pointer.}
We let each node $x$ maintain the $x[k]\dptr$ \textbf{only} for $k \ge c \triangleq x.\lci$ in a hash map \code{pointers}. 
This is sufficient for hash pointer reads, which happens only when a node $x$ receives a sequence of entries $E_{i:j}$ to be updated to its logs, plus the preceding pointer $E_{i-1}\dptr$. 
We may presume $j > c$ because $x$ rejects updating any committed entry. 
Normally, $x$ tells whether $E_{i:j}$ matches its own log list by whether $x[i-1]\dptr = E_{i-1}\dptr$. 
If $i \le c$, $x$ cannot find $x[i-1]\dptr$ in the hash map, but $x$ can alternatively derive $E_{c}\dptr$ by \eqref{eqn:hashptr} and tell whether $x[c]\dptr = E_{c}\dptr$. 
Since \method{Hash} is collision-resistant, $x[i-1]\dptr = E_{i-1}\dptr$ is implied by $x[c]\dptr = E_{c}\dptr$. 
Therefore, reduction of committed hash pointers (except the last one) does not affect correctness.



\textbf{Proposer Stamp.}
Suppose $\ell$ has proposed $\{E_k|k \in [i, j]\}$ during a leadership period. 
Since \FHash is collision resistant, $E_j\dptr$ effectively represents the entire log list from the head $E_1$ to $E_j$. 
Therefore, the stamp $\sigma_\ell(E_j\dptr)$ proves $\ell$ has proposed not only $E_j$, but also $E_{i}, \cdots, E_{j-1}$. 
This implies that the stamps on $E_{i}, \cdots, E_{j-1}$ are all redundant, and it suffices to keep only the last stamp $E_j$, e.g.,  
in a hash map \code{stamps} keyed by term. 

\textbf{Leader Certificate.}
By design, the \lcert used for each term is unique. 
Hence, we may reduce overheads by maintaining the \lcerts in a hash map \code{\lcmap} keyed by term and valued by \lcert. 
Moreover, we may reduce the hash pointer inside the vote request of \lcert, because the pointer can be derived from the logs. 

\textbf{Summary of Total Spatial Overhead.} 
Let $H$ denote the length of the logs, $H'$ the number of uncommitted entries and $\Lambda$ the number of global leaderships during which at least one entry is replicated.
Our optimized \sysname substantially reduces total overhead of the three states from $\mathcal O(nH)$ to $\mathcal O(H'+n\Lambda)$. 
However, to reduce notations and symbols for better clarity, we continue using the primitive states in the algorithm pseudocode. 


\subsection{Cost Analysis}
\label{sec:comparison-peerreview}

Using this optimized implementation, for Raft and \multipaxos, we compare the overhead space and communication complexities of  \sysname against PeerReview. 
For log replication, Raft is identical to \multipaxos, so we  merge the comparison in \S\ref{sec:perf-lr}. 
For leader election,  we compare the variants separately in \S\ref{sec:perf-le}. 

\paragraph*{PeerReview}

PeerReview \cite{peerreview} achieves accountability by logging communication for every message from any node $x$ to another node $j$, regardless of the underlying consensus protocol. 
The communication log is an independent data structure introduced by PeerReview. 
We call such log entries ``comm entries'', where each comm entry includes a copy of the message. 
To make the entire log tamper-evident, a hash pointer is maintained, just as in \sysname. 
We assume each comm entry stores a hash pointer, though this storage cost can be reduced by storing one pointer every few blocks, at the expense of time complexity of random access.
For every message \code{msg} sent from $x$ to $y$, $x$ sends \code{msg} along with a hash pointer and $x$'s signature. 
Then, $y$ replies a hash pointer plus $y$'s signature to $x$. 
Both $x$ and $y$ create a new comm entry including a copy of \code{msg}.
Hence, each message incurs communication overheads of two hash pointers and two signatures.

For auditing, PeerReview allows nodes to supervise each other by forwarding all signatures from a signer to the signer's \emph{witnesses}.
For a fair comparison between \sysname (which has a separate auditor) and PeerReview, we disable witnessing.

\subsubsection{Log Replication}
\label{sec:perf-lr}

\begin{table*}[!htb]
    \centering \small
    \begin{tabular}{c|c|*{2}{c}}
    \hline
         & Raft/Paxos & \textbf{\sysname (ours)} & PeerReview  \\ 
    \hline
         & (Base) & \multicolumn{2}{c}{Communication Overhead}\\ 
    \hline
        \method{Heartbeat} & \textsc{const} & $0$ & $ 2(\Pi+\Sigma)$ \\
        \method{AppendEntries} & $mB$ & $\Pi+2\Sigma$ \hide{$\Sigma + [\tau+1 + (\tau-1)(n-f)]\Sigma$} & $4(\Pi+\Sigma)$  \\
        \method{InformCommit} & \textsc{const} & $\Pi+(n-f)\Sigma$ & $0$  \\
    \hline
         & (Base) & \multicolumn{2}{c}{Storage Overhead}\\ 
    \hline
        \method{Heartbeat} & 0 & 0 & $2(\Pi+\Sigma)$  \\
        \method{AppendEntries} & $mB$ & 0\hide{$(\tau-1)(n-f+1)\Sigma$} & $2mB + 4(\Pi+\Sigma)$  \\
        \method{InformCommit} & 0 & 0 & $0$  \\
    \hline
    \end{tabular}
    \caption{Complexities of Raft/\multipaxos, \sysname and PeerReview in log replication.  $\Pi$ denotes hash size and $\Sigma$ denotes digital signature size, both in bytes. $m$ denotes number of log messages.}
    \label{tab:perf-lr}
    \vspace{-0.2in}
\end{table*}

\begin{figure}[!htb]
    \centering
    \includegraphics[width=\linewidth]{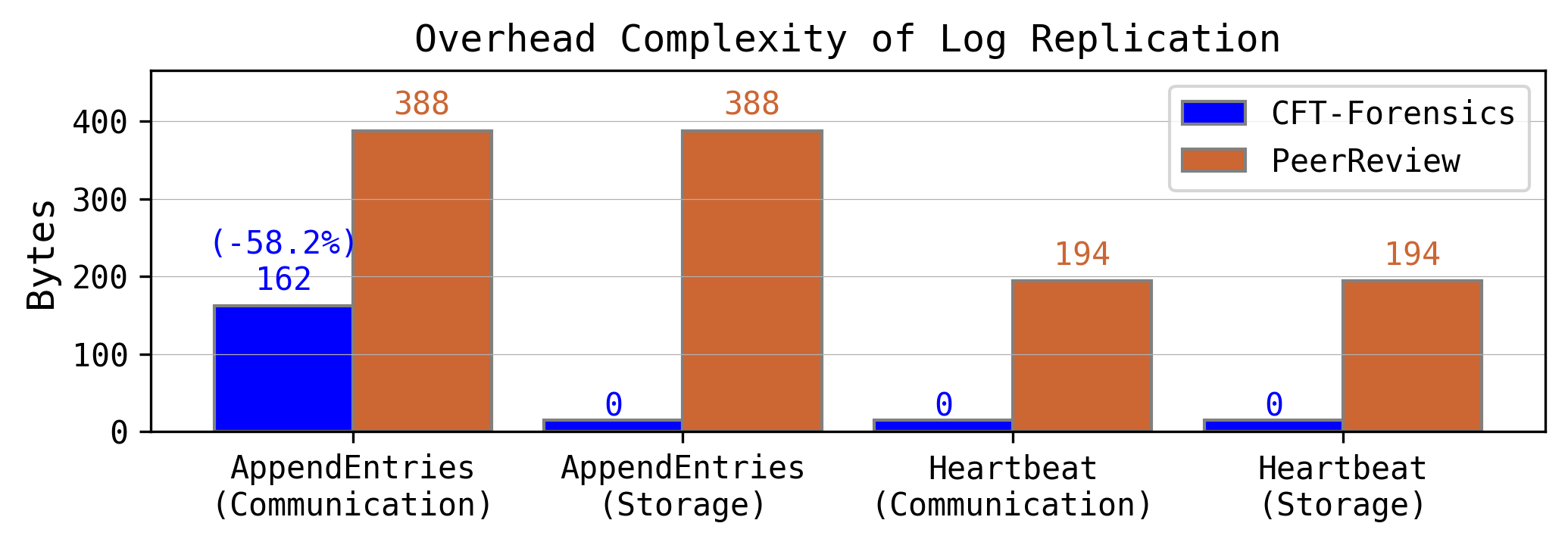}
    \caption{Overhead complexities of \sysname and PeerReview in log replication when hash size $\Pi = 32$ bytes and digital signatures are $\Sigma = 65$ bytes. }
    \label{fig:perf-lr}
\end{figure}


Let $\Pi$ and $\Sigma$ denote the sizes of a hash and a digital signature, respectively. 
We choose $\Pi = 32$ bytes and $\Sigma = 65$ bytes for numerical estimation, which are used for Ethereum\cite{ethyellow}. 
Let $B$ denote the size of a log entry. 
For messages including a sequence of log entries, we let $m$ denote the number of entries.  
We assume nodes are up-to-date in term and need onlyreplicate entries of current term. 
This limits the number of stamps and \lcerts sent along with the sequence. 
We also assume that \method{AppendEntries} complete in a single round, 
and that \method{InformCommit} contributes negligible overhead
(App. \ref{sec:lr-setting}). 

Table \ref{tab:perf-lr} presents the communication and storage complexities of Raft/\multipaxos, \sysname and PeerReview in three main log replication RPCs. 
For our assumed parameter values, we numerically visualize  the overheads of the \method{Heartbeat} and the \method{AppendEntries} RPCs in Fig. \ref{fig:perf-lr}. 
We first observe that \sysname has \textbf{zero} storage overhead in all three RPCs, while PeerReview has a positive overhead for \method{Heartbeat} and \method{AppendEntries}. 
Since message frequency must be lower-bounded by the \method{Heartbeat} frequency which is typically once every several seconds, 
\sysname outperforms PeerReview by saving about 1 KB storage every minute. 
For communication complexity, we focus on the  most frequently-used RPC:  (one-round) \method{AppendEntries}. 
\sysname has a $(\Pi+2\Sigma=162)$-byte overhead in communication, which is 58.2\% lower than $4(\Pi+\Sigma) = 388$ bytes of PeerReview. 





\subsubsection{Leader Election}
\label{sec:perf-le}

\begin{table*}[!htb]
    \centering \small
    \begin{tabular}{cc|c|*{2}{p{12em}<{\centering}}}
    \hline
         & & Vanilla (base) & \textbf{\sysname (ours)} (Overhead) & PeerReview (Overhead) \\ 
    \hline
        \multirow{2}{*}{Raft} & Comm. & \textsc{const} & $\Pi+(n-f)\Sigma$ & $ 6(\Pi+\Sigma)$  \\
        & Storage & 0 & $(n-f)\Sigma$ & $6(\Pi+\Sigma)$ \\
    \hline
        \multirow{2}{*}{\Multipaxos} & Comm. & $mB$ & $\Pi + (n-f+1)\Sigma$ \hide{$\Pi+[\tau+(\tau+I)(n-f)]\Sigma$} & $4(\Pi+\Sigma)$  \\
        & Storage & 0 & $\tau(n-f)\Sigma$ & $2mB+4(\Pi+\Sigma)$ \\
    \hline
    \end{tabular}
    \caption{Comparison of \emph{overhead} complexities between \sysname and PeerReview in leader election. 
    $I = 0$ if the candidate's last committed entry is at the same term as the first entry it receives from the voter; $I = 1$ otherwise.
    If a voter contributed a signature to the new \lcert, the \lcert it receives from the candidate does not need to include its own signature. 
    }
    \label{tab:perf-le}
    \vspace{-0.2in}
\end{table*}

\begin{figure}[!htb]
    \centering
    \includegraphics[width=\linewidth]{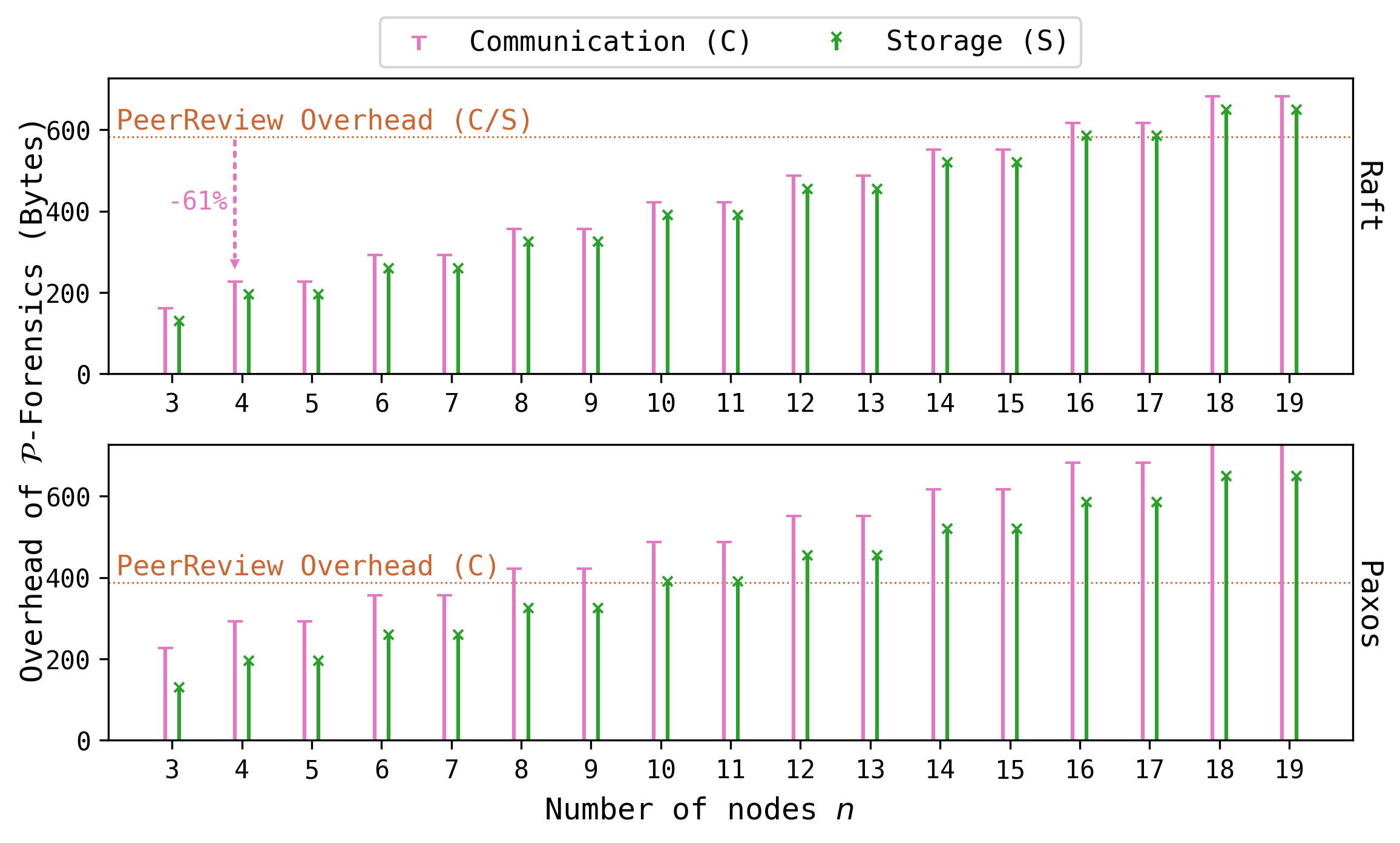}
    \caption{Overhead complexities of \sysname and PeerReview in leader election.  
    }
    \label{fig:perf-le}
    \vspace{-0.1in}
\end{figure}

\paragraph*{\raftname vs Raft-PeerReview}

Now we consider Raft's leader election, as described in Alg. \ref{alg:raft-impl}. 
A successful election has three messages between a candidate $\ell$ and its voter $x$: 1) $\ell$ sends vote request to $x$; 2) $x$ responds with a vote; and 3) $\ell$ sends a leadership claim. 
As shown in Table \ref{tab:perf-le} and Fig. \ref{fig:perf-le}, although \lcert contributes an $\mathcal O(n)$ overhead to \sysname, both complexities are still lower than Raft-PeerReview for $n \le 15$ {(under our assumed parameter values)}.  

\paragraph*{\paxosname vs Paxos-PeerReview}

A successful \multipaxos leader election (Alg. \ref{alg:paxos-impl})
has two messages between a candidate $\ell$ and its voter $x$: 1) $\ell$ sends its \lci to $x$; 2) $x$ responds with all its entries starting with $\lci+1$. 
In \paxosname, we insert three more messages: 3) $\ell$ sends a vote request to $x$; 4) $x$ responds with a signed vote and 5) $\ell$ sends a \lcert to claim leadership. 
Table \ref{tab:perf-le} lists the overheads for \multipaxos. 
We assume that leader elections are rare, 
so message 2) only includes entries of same term as $\ell[\lci]$. 
By Fig. \ref{fig:perf-le}, \multipaxos has lower communication complexity than Paxos-PeerReview if $n \le 7$, 
{and on a long enough timescale, its storage complexity is arbitrarily lower than that of Paxos-PeerReview.}

\section{Empirical Evaluation}
\label{sec:eval}

We implement  \raftname\footnote{\redact{\url{https://github.com/proy-11/NuRaft-Forensics.git}}}
 in C++ based on nuRaft v1.3 \cite{nuraft} by eBay. 
With roughly 2,500 lines of code, our implementation fully expands nuRaft with our OpenSSL-based designs in log replication, which correctly reflects the throughput and latency performances {between leader elections}.
We choose the SHA-256 hash function and Elliptic Curve Digital Signature Algorithm (ECDSA) over the secp256r1 curve. 
For commitment certificates, we used concatenated ECDSA signatures by all the signers.

We evaluate \raftname in two phases -- online phase (\sref{sec:eval-online}) and offline phase (\sref{sec:eval-offline}). 
In the online phase, we benchmark the performance of \raftname over a WAN.
In the offline phase, we evaluate the auditing procedure that scans server logs for adversarial behaviors. 

\subsection{Online Evaluation}
\label{sec:eval-online}



\paragraph*{Setup on AWS}
{We evaluate \raftname over a WAN to demonstrate a geo-redundant deployment for increased resilience \cite{georedundancy}.}
We simulated the WAN environment by deploying \raftname and other baseline protocols on multiple \texttt{c5.large} instances on AWS, 
where each instance has 2 vCPUs and 4 GB Memory. 
{
    We ran the experiments on 4 and 16 instances, respectively. 
    Because some typical applications of \raftname require the nodes to be distributed domestically, we deployed the 16 instances evenly in 8 AWS datacenters in the US, Canada and Europe. 
    For the 4-instance experiments, we deployed the instances in 4 US datacenters. 
}

\paragraph*{Baseline Protocols}
We compare the performance of \raftname against Raft \cite{nuraft}, 
using eBay's NuRaft \cite{nuraft} implementation. 
{We do not directly compare to state-of-the-art BFT protocols in our evaluation because our goal is to propose low-cost solutions that can be easily integrated into existing systems (i.e., the implementation should build upon existing code, and hence be some variant of Raft). 
Although there exist BFT variants of Raft  \cite{t1,t2,t3}, we were unable to confirm essential theoretical details needed to understand the protocol and guarantees. 
For completeness, we compare \raftname against a recent BFT protocol called Dumbo-NG \cite{gao2022dumbo} in Appendix \ref{sec:discussion-bft},
though a fair comparison is challenging and not the focus of this work. 
}

\paragraph*{Experimental Settings}
We benchmark each protocol by two metrics -- transaction latency and throughput. 
Latency is measured by the average time difference between when a transaction is confirmed and when it is sent to the servers. 
Throughput is measured by the average number of transactions processed per second during an experiment. 

\begin{figure}[!htb]
    \centering
    \includegraphics[width=\linewidth]{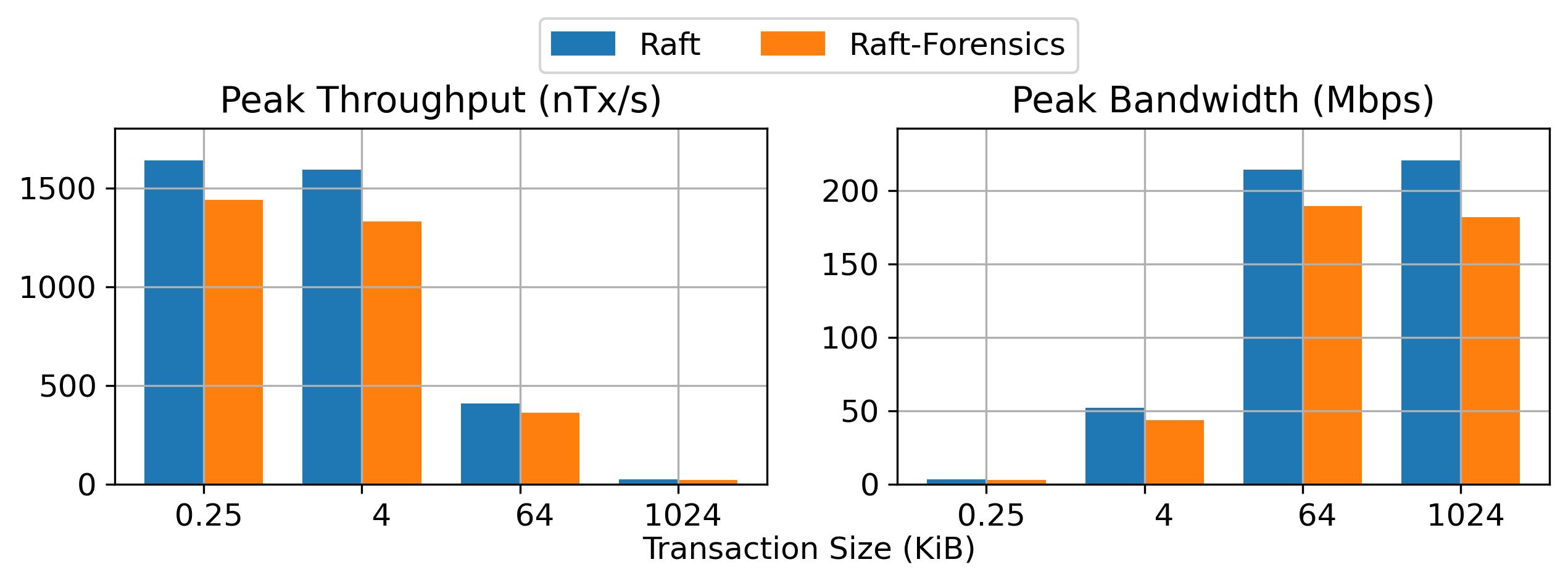}
    \caption{Peak transaction and bandwidth throughputs of consensus algorithms. ($n = 4$ nodes)}
    \label{fig:peaktpt}
\end{figure}

\begin{figure*}[!htb]
    \centering
    \includegraphics[width=\linewidth]{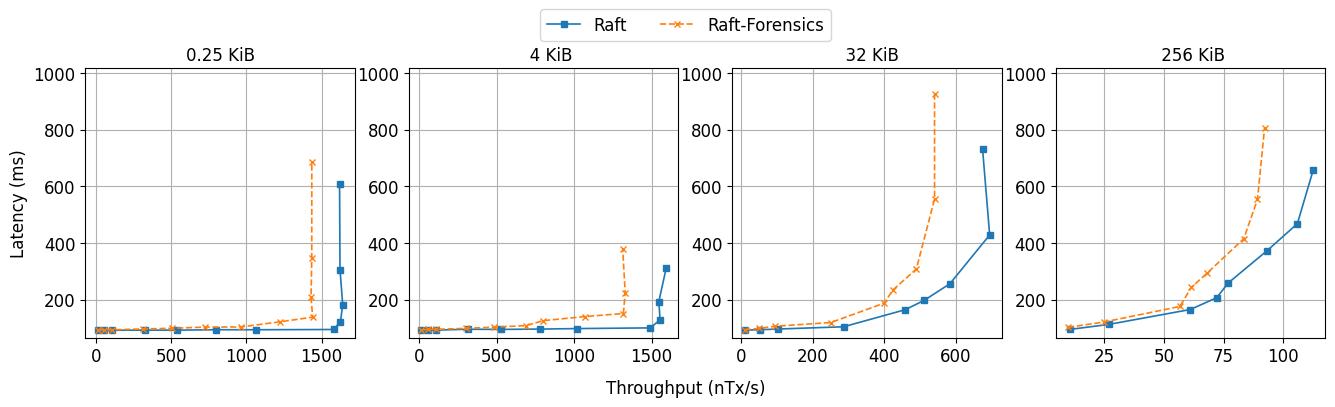}
    \includegraphics[width=\linewidth]{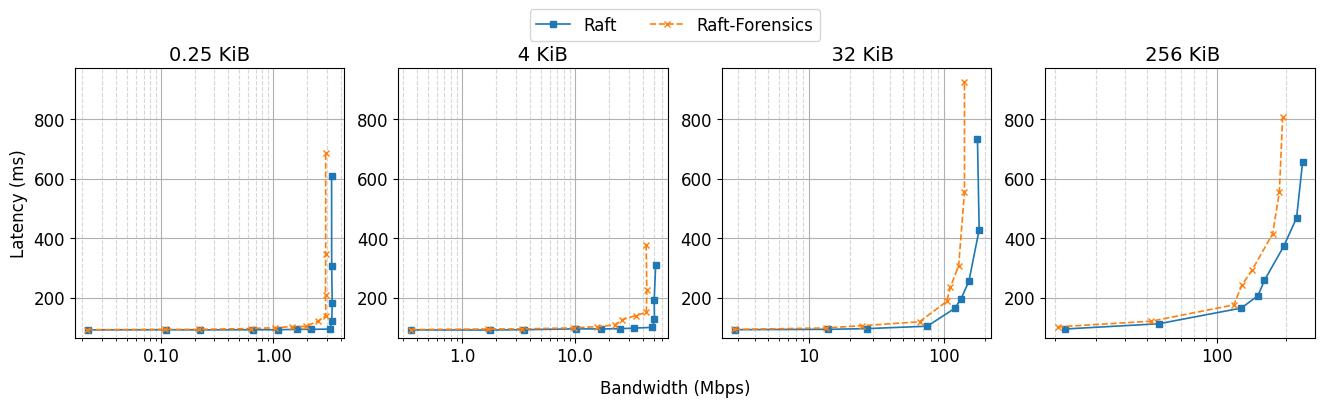}
    \caption{Latency-throughput tradeoff ($n = 4$ nodes). Top row displays throughput in number of transactions per second; bottom row displays throughput in bandwidth. 
    }
    \label{fig:lat}
\end{figure*}

The experiments are configured by two key parameters -- transaction size and number of concurrent clients. 
The transaction sizes range from 256 Bytes to 1 MB. 
For each transaction size, we {sweep the number of concurrent clients sending transactions {(in experiments, we let the leader machine spawn transactions)}. }
Under each configuration {of transaction size and client concurrency}, 
we run all the nodes and client processes simultaneously for 20 seconds.
We measure transaction latency and throughput by the average of five repeated runs to reduce random perturbations.
Typically, as the number of clients increases, throughput increases first linearly and then plateaus when the protocol is saturated. 
In contrast, latency is insensitive to the number of clients before the saturation, but rapidly increases when the bottleneck throughput is reached. 
We finally evaluate the following quantities:
\begin{itemize}
    \item \textbf{Peak throughput.} We measure the peak throughput of each baseline as the maximum number of transactions processed per second over all numbers of concurrent clients. 
    \fref{fig:peaktpt} presents the performance of all protocols under transaction sizes of 256 Bytes, 4 KiB, and 64 KiB. 
    Compared to Raft, \raftname has an approximately 10\% loss in peak throughput under various transaction sizes, which is caused by the cryptographic operations involved. 

    \item \textbf{Latency-Throughput tradeoff.} 
    Under each transaction size, we measure the latency-throughput curve parameterized by number of concurrent clients.   
    \fref{fig:lat} shows the latency-throughput tradeoffs of the two protocols under various transaction sizes. 
    Generally, the tradeoff of \raftname is only slightly worse than Raft. 
\end{itemize}

\subsection{Offline Evaluation}
\label{sec:eval-offline}

We next evaluate the offline performance of log auditing. 
Theorem \ref{thm:cft-acc} ensures that we can find at least 1 culprit when State Machine Safety is violated,
and we further show \emph{how} the culprit is found by Alg. \ref{alg:audition} and \ref{alg:audition-early} in Appendix \ref{sec:audit}.
Because the algorithm requires validity of nodes' data, Alg. \ref{alg:integrity}  (Appendix \ref{sec:chain-integrity})  checks data integrity before the audit. 

\paragraph*{Complexity Analysis}
Recall that $n$ denotes the number of nodes. 
Let $H$ denote the length of the longest chain and $\Lambda$ the number of elections in total. 
Table \ref{tab:audit-complexity} in Appendix \ref{sec:audit} summarizes the computational complexity of different parts of the auditing process. 
The total time complexity of auditing is asymptotically optimal (linear in  the size of data $n(H+\Lambda)$, which is required at minimum to ensure data legitimacy), 
where the complexity of global consistency checks does not depend on the chain length $H$. 
The linear spatial complexity $\Theta(n(H+\Lambda))$ requires chunked storage of the log chain. 
For instance, for a chunk size $\Theta(\log H)$, the spatial complexity decreases to $\Theta(n(\log H + \Lambda))$, while the time complexity remains the same. 
Notably, the time complexity of global consistency check slightly increases to $\mathcal{O}(n(\Lambda + \log^2 H))$, but is still much less than that of legitimacy checks. 

\paragraph*{Implementation }
We implement the auditing algorithm in Python\footnote{\redact{\url{https://github.com/WeizhaoT/Raft-Forensics-Simulator}}\label{footnote:sim}}, 
which can be tested along with a lightweight Raft simulator that achieves better control that the fully-implemented \raftname in C++ over the leader elections, the adversarial nodes' behavior and race conditions in general. 
In particular, it is capable of assigning the adversary to a node and simulating the fork and bad vote attacks in Examples \ref{eg:log-manip} and \ref{eg:obs-voting}. 
It ensures that the adversary generates legitimate data to prevent it from being caught before consistency checks. 
For the best performance in memory usage, it writes the data into chunked files that are available for auditing. 
    In Appendix \ref{sec:eval-offline}, we run benchmarks on the performance of both the data legitimacy and consistency checks of the auditing algorithm. 
    The benchmarks are consistent with our complexity analysis, and demonstrate a significant advantage in chunking data.  

Based on the backend software above, we also implement a visualizer based on \cite{DiemForensics} that demonstrates the attacks and the outputs of the auditing algorithm, including the identity of the culprit and the irrefutable evidence. 
Fig. \ref{fig:demo} shows a screenshot of the visualizer.

\subsection{Integration with OpenCBDC}
\label{sec:eval-cbdc}

Finally, we evaluate the performance of \raftname integrated into a downstream application: OpenCBDC \cite{lovejoyhamilton}, an open-source implementation of a retail central bank digital currency. 
OpenCBDC is a good choice because (a) it uses nuRaft, and (b) CBDCs are/will be public infrastructure, so security and performance are paramount.  
After integrating our \raftname implementation into OpenCBDC, we deployed our experiments onto \texttt{c5n.9xlarge} ec2 instances in AWS over three regions: \texttt{us-east-1}, \texttt{us-east-2} and \texttt{us-west-2}.\footnote{Although CFT protocols are often run in the same datacenter, if they are used for critical infrastructure, there will be a need for geographically-distributed deployments for robustness reasons.}

We compared \raftname against Raft in two different OpenCBDC architectures -- two-phase-commit (2pc) and atomizer. 
In the 2pc architecture, we created one generator, one sentinel, three coordinators and three shards, where each coordinator and each shard are Raft-replicated, i.e., they are implemented as Raft-variant distributed systems. 
In the atomizer architecture, we created one watchtower, one watchtower CLI, one sentinel, one archiver, four shards and three atomizers, where only atomizers are Raft-replicated.  
In both architectures, each Raft-replicated module consists of 3 nodes in 3 different AWS regions. 

We used the benchmarking platform \cite{opencbdc-tctl} of OpenCBDC under default configurations, 
where load generators produce as much workload as the system can process. 
The transaction size is 368 bytes. 
Each experiment lasts 5.25 minutes and is repeated 3 times. 
Table \ref{tab:result-cbdc} shows the throughput and latency of transactions of the entire system. 
We observe that in practical complex systems like OpenCBDC, \raftname also performs close to Raft.  

\begin{table}
    \small
    \centering
    \begin{tabular}{ccc}
    \hline 
        & \textbf{Throughput (\# tx/s)} & \textbf{Latency (ms)} \\
    \hline    
        \multicolumn{3}{c}{\textbf{2pc architecture}}  \\
    \hline 
        Raft & $4,800 \pm 14$ & $2,251 \pm 70$ \\
        \raftname & $4,695 \pm 76$ & $2,577 \pm 252$ \\
        (\% Change) & -105 (-2.2\%) & +326 (+14.5\%) \\
    \hline 
        \multicolumn{3}{c}{\textbf{atomizer architecture}}  \\
    \hline 
        Raft & $1,284 \pm 56$ & $37,552 \pm 18,75$ \\
        \raftname & $1,250 \pm 123$ & $40,802 \pm 1,653$ \\
        (Change) & -34 (-2.6\%) & +3,250 (+8.7\%) \\
    \hline 
    \end{tabular}
    \caption{Throughput and latency of two different OpenCBDC architectures integrated with Raft and \raftname (ours), respectively. 
    Each entry is expressed in $\text{mean} \pm \text{std}$.
    \vspace{-3ex}
    }
    \label{tab:result-cbdc}
\end{table}

\section{Discussion and Conclusion}
\label{sec:discussion}


This work is driven by the motivation to improve the Byzantine resistance of CFT protocols by introducing accountability, without sacrificing too much performance. One alternative approach to achieving higher security assurances with CFT protocols involves employing BFT protocols directly. This strategy not only increases tolerance to Byzantine faults but may also inherently include accountability as a bonus feature. 

As explained in Section \ref{sec:eval}, we were unable to directly compare against BFT variants of Raft \cite{t1,t2,t3}. 
Hence, we conducted performance comparisons between Raft-Forensics and leading BFT designs like Dumbo-NG, as detailed in Appendix \ref{sec:discussion-bft}. Our analysis indicates that, in terms of reducing latency, Raft-Forensics generally surpasses Dumbo-NG, though the latter may display competitive or superior throughput for larger transaction volumes. Moreover, Dumbo-NG is optimized for efficiently propagating blocks containing multiple transactions among numerous participants, while Raft variants typically handle single-transaction blocks (as required by SMR) in small-scale distributed systems. As a result, we acknowledge that BFT protocols can indeed be optimized to achieve good performance and replace CFT protocols in applications requiring higher security guarantees, albeit at the cost of increased design complexity and an overhaul of the entire consensus logic. In contrast, accountability may be more suitable for scenarios with moderate security improvement requirements and an emphasis on lightweight changes.

More broadly, accountability need not be viewed as an alternative to Byzantine fault tolerance---it is a complementary, desirable property. For example, all BFT protocols do not inherently offer accountability \cite{sheng2021bft}. 
We posit that accountability is an important component of distributed system governance---all the more so for geographically-distributed critical infrastructure \cite{ezzeldin2021robustness}.

\Urlmuskip=0mu plus 1mu\relax

\bibliography{bibfile}

\clearpage

\appendix


\section{Raft and \Multipaxos Implementation}

\label{sec:raft-paxos-code}

\begin{algorithm}[!htb]
\caption{Raft Implementation. 
\fmod{The red lines and phrases are specific to CFT-forensics (\S\ref{sec:cftforensics})}. 
}
\label{alg:raft-impl}

\Pn{\FLR{host node $z$}}{
    \textbf{As Leader:}\;
    \Rn{\method{HandleAEFollowUp}($x$, $i$)} {
        Call RPC \method{AppendEntries}($x$, $z[i:\mend]$)\;
    }
    
    \textbf{As Follower:}\;
    \Rn{\method{HandleAppendEntries}($\ell$, $E_{i-1:j}$)} {
        \If{$E_{i-1} = z[i-1]$  } {
            \If(\tcp*[f]{Check conflict with committed entries}){$\exists k \in [i, \lci]: E_k \neq z[k]$  } {
                \textbf{fail exit}\;
            }
            \color{modclr}
            \If{\KwNot $\method{validateNewState}(E_{i-1:j})$} {
                \textbf{fail exit}\;
            }
            \color{black}
            
            $z[i:\mend] \leftarrow E_{i:j}$\;
            Send \codep{AppendEntriesResp}($E_j$, \fmod{$\sigma_z(E_j\dptr)$}) to $\ell$\;
        }\Else {
            Call RPC \method{AEFollowUp}($\ell$, \lci)\; \label{line:raft-sync}
        }
    }
    \Rn{\method{HandleInformCommitMsg}($\ell$, \msg)} {
        \If{$z$ owns \texttt{msg.entry} \fmod{ \KwAnd $\mathtt{validate}(\msg.\ccert)$ }} {
            \color{modclr}
            $\ccert \leftarrow \msg.\ccert$\;
            \color{black}
            
            $\method{Commit}(\texttt{msg.entry})$\;
        }\ElseIf{$E_i \neq z[i]$} {
            Call RPC \codep{AEFollowUp}($\ell$, \lci)\;
            \method{HandleInformCommitMsg}($\ell$, \msg)\;
        }
    }
}

\Pn{\FLE{host node $z$}}{
    
    \Fn{\method{Qualification}(\texttt{term})} {
        
        $\texttt{votedFor}, \texttt{voters} \leftarrow z, \{z\}$\;
        
        $r \leftarrow z \| \mathtt{term} \| z[\mend]\dterm \| \mend \color{modclr} \| z[\mend]\dptr \color{black}$\;
        
        \color{modclr}
        $\mathtt{sigs} \leftarrow \{ \sigma_z(\mathtt{Hash}(r)) \}$ 
        \color{black}

        \For{async $x \in S - \{z\}$ } {
            \msg $\leftarrow$ call RPC \codep{RaftRequestVote}($r$)\;
            
            \color{modclr}
            \If{\KwNot \texttt{verifySig}(\texttt{msg.signature})} {
                \textbf{Abort} $x$\;
            }
            \color{black}
            
            $\mathtt{voters},~ \fmod{\mathtt{sigs}} \leftarrow \mathtt{voters} \cup x, ~ \fmod{\mathtt{sigs} \cup \{\texttt{msg.signature}\}}$\;
        }

        Wait until $|\texttt{voters} | \ge n-f-1$\;
        
        \color{modclr}
        $\lcert \leftarrow r \| \mathtt{voters} \| \mathtt{sigs} $\;
        \color{black}

        \KwRet \True\;
    }
    \Rn{\method{HandleRaftRequestVote}($\ell$, \texttt{req})} {
        \If{$\ell\dterm > \texttt{term}$} {
            Interrupt \codep{CandidateMain}\;
            $\texttt{term}, \texttt{voters} \leftarrow \ell\dterm, \emptyset$\;
            
            \If{$z[\mend]\dterm < \mathtt{req.eterm}$ \Or $z[\mend]\dterm = \mathtt{req.eterm} \wedge z[\mend]\dindex \le \mathtt{req.end}$} {
                $\texttt{votedFor} \leftarrow \ell$\;
                Send \codep{Message}(\texttt{accept}=\True, \fmod{\texttt{signature}=$\sigma_z(\texttt{Hash}(\texttt{req})$}) to $\ell$\;
            } \Else {
                Send \codep{Message}(\texttt{accept}=\False) to $\ell$\;
            }
        }
    }
}
\end{algorithm}

\begin{algorithm}[!htb]
\caption{RPC implementations of \Multipaxos. The log replication RPCs are identical to Raft in Alg. \ref{alg:raft-impl}.
\fmod{The red lines and phrases are specific to CFT-forensics (\S\ref{sec:cftforensics})}. }
\label{alg:paxos-impl}

\Pn{\FLE{host node $z$}}{
    \textbf{As Candidate:}\;
    \Fn{\method{Qualification}(\texttt{term})} {
        $\mathtt{pusher}, \texttt{lg}, m \leftarrow \emptyset, \texttt{logList.clone()}, \mend$\;

        \For{async $x \in S - \{z\}$} {
            \msg $\leftarrow$ Call RPC \method{PullLog}(\texttt{term}, \lci)\;
            
            \color{modclr}
            \If{\KwNot \method{validateNewState}($\msg\texttt{.logs}$)} {
                \textbf{Abort} $x$\;
            }
            \color{black}
        
            \texttt{mutex.acquire()}\;

            $m \leftarrow \max\{m, \msg\texttt{.end}\}$\;
            \For{$k \in \mathbb Z \cap [\lci+1, m]$} {
                \If{$\texttt{lg}[k] = \perp$ \Or $\texttt{lg}[k]\dterm < \msg\texttt{.logs}[k]\dterm$} {
                    $\texttt{lg}[k] \leftarrow \msg\texttt{.logs}[k]$\;
                    \color{modclr}
                    $\texttt{lg}[k]\dptr \leftarrow \mathtt{Hash}(k\| \texttt{lg}[k]\dpay \| \texttt{lg}[k-1]\dptr)$\;
                    $\texttt{lg}[k]\texttt{.signature} \leftarrow \sigma_z(\texttt{lg}[k]\dptr)$\;
                    \color{black}
                }
                $\texttt{lg}[k]\dterm \leftarrow \texttt{term}$\;
            }
            $\mathtt{pusher} \leftarrow \mathtt{pusher} \cup \{x\}$\;
            \texttt{mutex.release()}\;
        }
        
        Wait for $|\mathtt{pusher}| \ge n-f-1$\;

        $z[\lci+1:m] \leftarrow \texttt{lg}[\lci+1:m]$\;

        \KwRet \True\;
    }
    \Rn{\method{HandlePullLog}($x$, $\tau$, $i$)}{
        \If{$\tau > \texttt{term}$}{        
            Send \codep{Message}(\texttt{end}=\mend, \texttt{logs}=$z[i:\mend]$) to $x$\;  
        }
    }
}
\end{algorithm}



\section{Proof of Thm. \ref{thm:cft-acc}}
\label{sec:proof-safety}


\textbf{Proof.}
We use the optimized states (\S\ref{sec:stateopt}) instead of basic states (\S\ref{sec:state}) for clarity in this proof. 
Recall that optimized states can be derived from basic states by redundancy reduction, so the correctness of this proof is not affected by how the states are maintained. 

When safety (Def. \ref{def:smr}) is breached, there exists two honest nodes $u$ and $v$ with conflicting {committed} entries. 
Namely, the term, index of $u$'s and $v$'s \textbf{last committed entries }are $(\tau_u, j_u)$ and $(\tau_v, j_v)$, respectively. 
By the protocol $\Pc$, if an honest node owns an entry at term $\tau$, it must own the \lcert of the leader at term $\tau$. 

We first check whether each term can be mapped to a unique leader. 

\textbf{Case 0.} 
There exists term $\tau$, such that $u.\lcmap[\tau]$ and $v.\lcmap[\tau]$ belong to different leaders. 
An auditor can accuse all the nodes in $u.\lcmap[\tau].\texttt{voters} \cap v.\lcmap[\tau].\texttt{voters}$ because they illegally voted twice at the same term (check \method{HandleClaimLeadershipMsg} in Alg. \ref{alg:elect-general}). 

For the next cases, we assume $\tau_u \ge \tau_v$ without loss of generality, and there are no \lcert-\lcert conflicts.

\textbf{Case 1 (against the split-brains attack).} There is a conflict within term $\tau_v$. 
In other words, $u$ and $v$ own two entries $\epsilon_u$ and $\epsilon_v$ respectively, where both entries are at term $\tau_v$, but neither is an ancestor or descendant of another. 
An auditor is able to accuse the leader of term $\tau_v$, 
where the evidences are $u.\code{stamps}[\tau_v]$ and $v.\code{stamps}[\tau_v]$, i.e., the proposer stamps on term $\tau_v$'s last entries, which are different for nodes $u$ and $v$.  

\textbf{Case 2 (against voter's attack).} There is no conflict within term $\tau_v$. 
We can assert that $\tau_u > \tau_v$, because otherwise we must have $\tau_u = \tau_v$ and this denies our very first assumption -- $u$ and $v$ have conflicting entries. 
Let $U \triangleq \{u[i]\dterm|i\in\mathbb Z_{>0}\}$ be the set of terms of $u$'s log entries. 
We pick $\tau \triangleq \min \{ \tau | \tau \in U,~ \tau > \tau_v \}$, which exists because $U$ has at least one element $\tau_u$. 
We consider term $\tau$'s leader $\ell$ and its freshness $(\tau_\ell, j_\ell)$ when it requested for vote. 
By the minimality of $\tau$, $\tau_\ell \le \tau \le \tau_v$. 
Because there exists a conflict between $u$'s chain and $v$'s chain, $(\tau_\ell, j_\ell)$ must be strictly staler than $(\tau_v, j_v)$. 
Otherwise, we must have $\tau_\ell = \tau_v$ and $j_\ell \ge j_v$. 
In conjunction with the case assumption ``no conflict within term $\tau_v$'', we deduce that $v$'s chain is a prefix of $u$'s chain, a contradiction. 

Therefore, $(\tau_\ell, j_\ell)$ is strictly staler than $(\tau_v, j_v)$.
In this case, no honest node can vote for $\ell$ \textbf{after} replicating the fresher log entry at $(\tau_v, j_v)$ which makes $\ell$ disqualified for freshness in Property \ref{prop:election}. 
On the other hand, no honest node can vote for $\ell$ \textbf{before} signing the log entry at $(\tau_v, j_v)$, either. 
This is because by replication property (Property \ref{prop:repl}), the entry came from an older term $\tau_v$ than the node's term $\tau$ and should have been rejected.  

Therefore, no honest node can sign both $v.\ccert$ and the $u.\lcmap[\tau]$, regardless of the temporal order. 
An auditor can acquire both certificates if $u$ and $v$ appear honest. 
Then, by the pigeonhole principle,  $I \triangleq u.\lcmap[\tau].\texttt{voters} \cap v.\ccert.\texttt{voters}$ is non-empty and an auditor can accuse all nodes in $I$.

To summarize, if two different nodes have valid and conflicting states, 
at least one adversarial node can be accused with irrefutable proof by an auditor with access to their states. \qeda

\section{The Auditing Algorithm}
\label{sec:audit}

We present the pairwise consistency checking algorithm in Alg. \ref{alg:audition} and an early-existing high-level auditing algorithm in Alg. \ref{alg:audition-early}. 
The inputs correspond to the maintained states defined in \S\ref{sec:state}, where the relevant fields can be found in Table \ref{tab:attr}. 
Their time complexities are listed in Table \ref{tab:audit-complexity}. 

\begin{algorithm}
\caption{The auditing algorithm for pairwise consistency checks. Time complexity $\mathcal{O}(n + \Lambda)$. }
\label{alg:audition}
\KwIn{Nodes $u$ and $v$}
\KwOut{If $u$ and $v$ are consistent, $\varnothing$; otherwise a non empty set of Byzantine nodes}

\tcc{$u, v$ must pass \FIntegrate checks in Alg. \ref{alg:integrity}}

\Pn{\FAudit{$u$, $v$}}{
    \For{$t \in u.\lcmap\dkeys \cap v.\lcmap\dkeys$} {
        \If{$u.\lcmap[t]\dreq\dleader \neq v.\lcmap[t]\dreq\dleader$} {
            \KwRet $u.\lcmap[t]\dvoter \cap v.\lcmap[t]\dvoter$\tcp*{Case 0}
        }
    }
    $i_v,~ i_u \leftarrow v.\ccert\dindex,~ u.\ccert\dindex$\;
    \If{ $v[\min\{ i_v, i_u \}] = u[\min\{ i_v, i_u \}]$ } {
        \KwRet $\varnothing$ \tcp*{Chains are consistent}
    }
    \If{ $v.\ccert\dterm = u.\ccert\dterm$ } {
        \KwRet $\{v.\lcmap[v.\ccert\dterm]\dreq\dleader\}$\tcp*{Case 1}
    }
    $h \leftarrow \argmax_{u, v} \{ u.\ccert\dterm,~ v.\ccert\dterm \}$\;
    $\ell \leftarrow \argmin_{u, v} \{ u.\ccert\dterm,~ v.\ccert\dterm \}$\;
    $\tau \leftarrow \min\{t | t \in h.\lcmap\dkeys,~ t > \ell.\ccert\dterm\}$\;
    \If{$\ell.\ccert\dterm \in h.\lcmap\dkeys$} {
        $j \leftarrow h.\lcmap[\tau]\dreq.\mend$\;
        \If{ $\ell.\ccert[j] \neq h.\ccert[j]$ } {
            \KwRet $\{\lcmap_\ell[\ell.\ccert\dterm]\dreq\dleader\}$\tcp*{Case 1}
        }
    }
    \KwRet $h.\lcmap[\tau]\dvoter \cap \ell.\ccert\dvoter$\tcp*{Cases 2}
}
\end{algorithm}

\begin{algorithm}
\caption{The early-exiting auditing algorithm of \sysname. A full version can be found in Alg. \ref{alg:audition-full}.}
\label{alg:audition-early}
\KwIn{Node set $S$}
\KwOut{Byzantine nodes $A \subset S$}

\Pn{\FAuditAll{$L$, $E$, $C$, $S$}}{
    \For{$v \in S$} {
        \If{\KwNot \FIntegrate{$v$} } {
            \KwRet $\{v\}$\;
        }
    }

    $w \leftarrow \argmax_{v \in V} v\dlen$\;
    \For{$v \in S - \{w\}$} {
        $A_v \leftarrow $ \FAudit{$v$, $w$}\;
        \If{$A_v \neq \varnothing$} {
            \KwRet $A_v$\;
        }
    }
    
    \KwRet $\varnothing$\;
}
\end{algorithm}

\subsection{Chain Integrity Algorithm}
\label{sec:chain-integrity}

For each node $v$, the audited data consists of 4 components: the log list $v$, the commitment certificate $v.\ccert$, the \lcert map $v.\lcmap$ and leader signatures $v.\code{stamps}$. 
We list the requirements for the data of a node $v$ to be \emph{legitimate}. 

\begin{enumerate}
    \item $v$ starts with a dummy entry \code{InitLog} and lasts with the corresponding entry of the commitment certificate $v.\ccert$. 

    \item For each entry in $v$,
    \begin{enumerate}
        \item The \code{index} field equals the number of its ancestors. 
        \item The \code{term} field is no less than that of its predecessor.
        \item There exists a leader certificate in $v.\lcmap$ keyed by \code{term}. 
        \item If $v$ is the last entry of its term, the corresponding {signature} in $v.\code{stamps}$ can be verified with its hash pointer and the leader's public key.  
    \end{enumerate}

    \item The signatures in commitment certificate $v.\ccert$ are correctly signed by the voters in $v.\ccert$. Numbers of voters and signatures should equal and be at least $f+1$. 

    \item For each leader certificate with term $t$ in $v.\lcmap$, 
    \begin{enumerate}
        \item The signatures are correctly signed by the voters. Numbers of voters and signatures should equal and be at least $f+1$. 
        \item If there exists at least one entry at term $t$ in $v$, the predecessor of first entry at term $t$ must have the same freshness as the certificate. 
        \item (Ensured by Alg. \ref{alg:integrity}) A log entry with term $t$ exists in $v$. 
    \end{enumerate}
\end{enumerate}


Alg. \ref{alg:integrity} describes the verification procedure of a node's data integrity. 
If the data is not legitimate, the node is considered Byzantine-faulty and will not participate in the pairwise consistency checks in Alg. \ref{alg:audition}.
In other words, all nodes in Alg. \ref{alg:audition} are guaranteed to have legitimate data. 
Note that it is possible that an honest node accepts a leader, but does not nodete any entry from it for various reasons.
For simplicity in Alg. \ref{alg:audition}, we remove all the leader certificates of terms of which no entries exist in the log list. 

\begin{algorithm}[!htb]
\small
\caption{Chain integrity check of \sysname}
\label{alg:integrity}
\def\failure{\code{fail}}
\KwIn{Node $v$}
\KwOut{A predicate indicating whether the data of $v$ is valid, an updated \lcert map $v.\lcmap$, and an auxiliary \lcert map $v.\lcmap'$}

\Pn{\FIntegrate{$v$}} {
    \tcc{Return false if any assertion fails}
    \tcc{Check log list}
    \ASSERT{$v[0] = \code{InitLog}$}\;

    $\code{AllTerms}, H \leftarrow \varnothing, $ length of $v$'s log list\;
    $\code{LCTerms} \leftarrow \method{SortedList}(\{\lcert\dreq\dterm | \lcert \in \lcmap.\mathtt{values}\})$\;
    $h, j \leftarrow \perp, 0$\;
    \For{$i = 1, 2, \cdots, H$} {
    
        \ASSERT{$v[i]\dindex = i$ \KwAnd $v[i]\dterm = \code{LCTerms}[j]$} \tcp*{Ensure index correctness}
        
        $t \leftarrow v[i]\dterm$\;
        $\code{AllTerms} \leftarrow \code{AllTerms} \cup \{t\}$\;
        \ASSERT{$t \ge v[i-1]\dterm$} \tcp*{Ensure non-decreasing terms}

        \ASSERT{$t \in v.\lcmap\dkeys$ \KwAnd $t \in v.\code{stamps}\dkeys$ } \tcp*{Ensure log creator is traceable}
        
        $h \leftarrow \method{Hash}(h \| t \| v[i]\dindex \| v[i]\dpay)$\;
        \If{$i = H$ \Or $t < v[i+1]\dterm$} {
            \ASSERT{\FVSig{$v.\code{stamps}[t], h, v.\lcmap[t]$}} \tcp*{Ensure authenticity}
            
            \If{$i < H$}{
                \ASSERT{$v[i+1]\dterm \in v.\lcmap\dkeys$} \tcp*{Ensure \lcert correctness}
            }
        }
        \If{$i = \lcmap[j]\dreq\dindex$}{
            \ASSERT{$h = v.\lcmap[j]\dreq\dptr$}\;
            $j \leftarrow j + 1 $\;
        }
    }
    \ASSERT{$j = $ length of \code{LCTerms}}\;
    \tcc{Check Commitment Certificate}
    \ASSERT{$v.\ccert\dptr = h$ \Or \KwNot \FVCert{$v.\ccert$}}
    \tcp*{Check \lcert map}
    $v.\lcmap' \leftarrow $ Empty hash map\;
    \For{$\code{term} \in v.\lcmap\dkeys$ } {
        \ASSERT{\FVCert{$v.\lcmap[\code{term}]$}} \tcp*{Ensure \lcert correctness}
        
        \If{$\code{term} \notin \code{AllTerms}$} {
            $v.\lcmap'[\code{term}] \leftarrow v.\lcmap[\code{term}]$\;
            delete $v.\lcmap[\code{term}]$\;
            \Continue\tcp*{Remove unused terms}
        }

    }
    
    \KwRet (\True, $v.\lcmap$, $v.\lcmap')$\;
}

\end{algorithm}

\subsection{The Full Auditing Algorithm}
\label{sec:audit-full}

In Alg. \ref{alg:audition-full}, we specify the full version of auditing algorithm which does not exit early after detection of one Byzantine node. 
It exhausts the available data and detects as many Byzantine nodes as possible. 
As long as a Byzantine node participates in forking the consensus, it will be exposed by this algorithm. 
Hence, in comparison with the early-exit version (Alg. \ref{alg:audition-early}), the full version is recommended, because unlike the former, it does not miss the Byzantine nodes whose accomplice is exposed earlier (e.g., with illegitimate data). 
However, we address that if multiple Byzantine nodes take part in the same attack (for example, voting for two different leaders at the same term), both full and early-exit variants of the auditing algorithm can find all these nodes at the same time. 

As expected, it suffers from a slightly higher complexity. 
Suppose the number of conflicting branches equals $\beta$. 
The time complexity of global consistency check (Table \ref{tab:audit-complexity}) will be raised from $n\Lambda$ to $n\beta \Lambda$. 
However, compared to the high complexity of global legitimacy check, which is linear to the size of the log list, this change is negligible. 

\begin{algorithm}[!htb]
\caption{The full auditing algorithm of \sysname.}
\label{alg:audition-full}
\KwIn{Node set $S$}
\KwOut{Byzantine-faulty nodes $A \subset S$}

\Pn{\FAuditAll{$S$}}{
    $A,~ V \leftarrow \varnothing,~ \varnothing$\;
    \For{$v \in S$} {
        \If{\FIntegrate{$v$} } {
            $V \leftarrow V \cup \{v\}$\;
        } \Else {
            $A \leftarrow A \cup \{v\}$\;
        }
    }

    \While{$|V| > 1$} {
        $U, ~w \leftarrow \varnothing,~ \argmax_{v \in V} v\dlen$\;
        \For{$v \in V - \{w\}$} {
            $A_v \leftarrow $\FAudit{$v, w$}\;
            $A \leftarrow A \cup A_v$\;
            \If{$A_v \neq \varnothing$} {
                $U \leftarrow U \cup \{v\}$\;
            }
        }
        $V \leftarrow U$\;
    }
    \KwRet $A$\;
}

\end{algorithm}

\begin{table}[!htb]
    \centering
    \def\Oc{\mathcal{O}}
    \begin{tabular}{cc}
    \hline \hline
        Item & Complexity \\
    \hline \hline
        Data Legitimacy Check (Alg. \ref{alg:integrity}) & $\Oc(H + \Lambda)$ \\
        \textbf{Global legitimacy Check} & $\Oc(n(H+\Lambda))$ \\
    \hline
        Pairwise consistency check (Alg. \ref{alg:audition}) & $\Oc(\Lambda)$ \\
        \textbf{Global consistency check (Alg. \ref{alg:audition-early})} & $\Oc(n \Lambda)$ \\
    \hline
        Total time & $\Oc(n(H + \Lambda))$ \\
    \hline 
        Total space (memory) & $\Theta(n(H + \Lambda))$ \\
    \hline \hline
    \end{tabular}
    \caption{Computational complexity of auditing. Recall $H$ is the length of log list and $\Lambda$ is the number of unique terms.}
    \label{tab:audit-complexity}
\end{table}

\section{Other Evaluation Results}

\subsection{Settings of Log Replication Cost Analysis}
\label{sec:lr-setting}

\textbf{Single-Round \method{AppendEntries}.}
In our comparison in \S\ref{sec:perf-lr}, we assume all \method{AppendEntries} calls are one-round. 
Note that a second round does not affect \sysname, but doubles the overheads of PeerReview; so our comparison shows the \emph{worst-case} advantage of \sysname. 
The one-round assumption comes from a strategic leader who tracks each follower's progress and avoids sending log entries that cannot be directly attached to a follower's log list. 
The leader needs sophisticated scheduling of \method{AppendEntries} for all the followers, which eventually allows one round \method{AppendEntries} and makes $m$ linearly related to the request frequency. 

\textbf{Insignificant \method{InformCommit} Overheads.}
In vanilla and PeerReview, \method{InformCommit} is not sent individually, but rather attached to the message in the next call of \method{AppendEntries} in practice. 
Hence, vanilla and PeerReview have zero overheads, while \sysname has a positive overhead of \ccert. 
However, in \raftname, the frequency of \ccert dissemination can be flexibly lower than that of \method{AppendEntries} calls in an actual implementation. 
So we recognize the overhead of commitment as a minor part compared to \method{AppendEntries}. 

\subsection{Comparison Between Raft and Raft-Forensics over 16 Nodes}
\label{sec:eval-n}

\begin{figure*}[!htb]
    \centering
    \includegraphics[width=\linewidth]{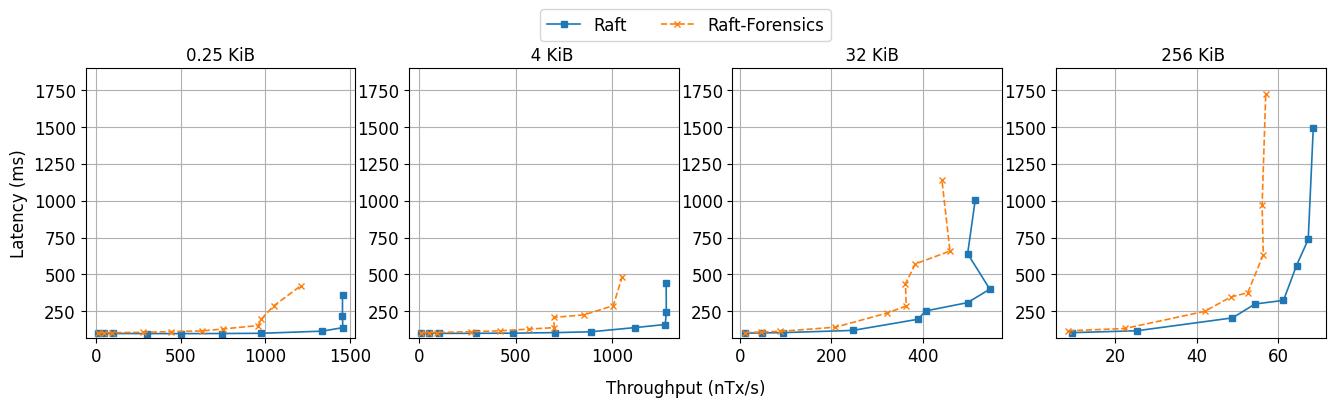}
    \includegraphics[width=\linewidth]{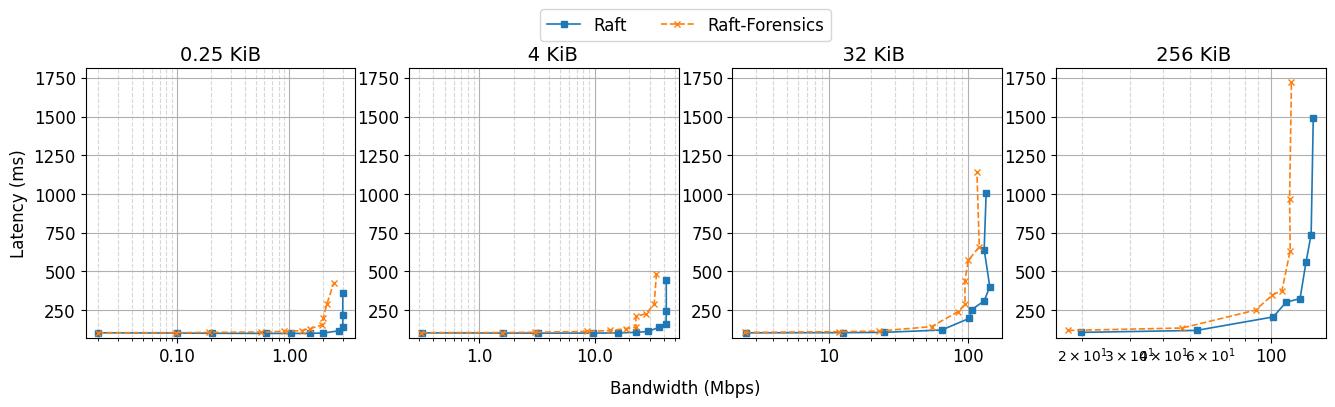}
    \caption{Latency-throughput tradeoff ($n = 16$ nodes). Top row displays throughput in number of transactions per second; bottom row displays throughput in bandwidth.}
    \label{fig:lat-16}
\end{figure*}

\subsection{Effect of Transaction Size}

We extend our evaluation results in \sref{sec:eval-online} by inspecting the effects of \emph{transaction size} on both latency and throughput. 
To observe these effects, we fix the client concurrency level and extend our choices for transaction sizes to a geometric sequence starting with 256 Bytes to 2 MiB. 
Each single experiment is configured by client concurrency and transaction size in the same way as in \sref{sec:eval-online}, and yields the throughput and latency of transactions. 
Because the concurrency levels in Dumbo-NG configurations are not comparable against Raft and \sysname, we only run the experiments for the latter. 

\begin{figure}[!htb]
    \centering
    \begin{subfigure}[b]{\linewidth}
        \includegraphics[width=\linewidth]{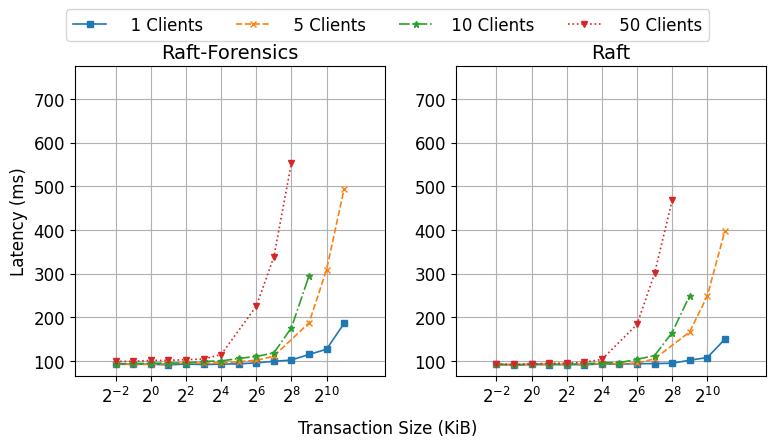}
        \caption{Relation between latency and transaction size.}
        \label{fig:lat-txsize}
    \end{subfigure}
    \begin{subfigure}[b]{\linewidth}
        \includegraphics[width=\linewidth]{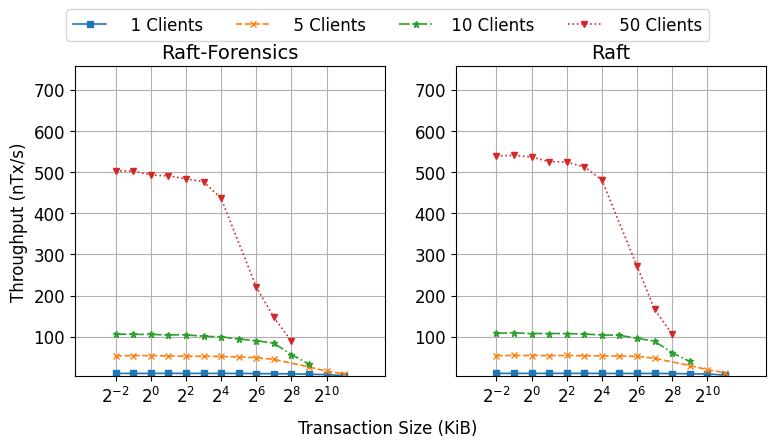}
        \caption{Relation between throughput and transaction size.}
        \label{fig:tpt-txsize}
    \end{subfigure}
    \caption{Relation between latency/throughput and transaction size. ($n = 4$ nodes)}
\end{figure}

Figures \ref{fig:lat-txsize} and \ref{fig:tpt-txsize} show the effects of transaction size on the system performance. 
In both figures, \sysname performs similarly to Raft -- over all experiments, its latency is at most 25\% higher than Raft, while its throughput is at least 93\% of Raft. 
When the number of concurrent clients is no greater than 10, the throughput of \sysname reaches 98\% of Raft at least, which makes them almost identical.
These figures are also helpful for system designers to choose a suitable transaction size to maximize bandwidth usage without sacrificing latency.
For instance, with 50 clients, 16 KiB is the most suitable transaction size, while with 10 clients, 128 KiB is optimal. 

\subsection{Performance of Auditing Algorithm}
\label{sec:audit-perf}

{In addition to theoretical complexity bounds, we simulate and benchmark the audit process.}
The audit process consist of two steps -- data generation and evaluation of auditing performance. 

To benchmark the audit process, we need to control:
1) the adversarial nodes' behavior,
2) the number of requests, which approximately equals the length of log chains, and
3) the chunk size of log chains. 
In our C++ implementation (as well as vanilla nuRaft) it is time-consuming and difficult to add different varieties of adversarial agents to generate desired test cases, particularly during elections (e.g., electing a leader that is favorable to the adversary).
Moreover, the full nuRaft implementation introduces a vast range of race conditions, which are challenging to control. 

As a result, for more fine-grained control over our audit experiments, we used the simulator to generate the data required for auditing. 
We summarize the functionalities of the simulator as follows.
\begin{itemize}
    \item It is able to fully simulate log replications with different pairwise node delays.

    \item It simplifies leader election so the choice of the new leader is controllable and suitable for testing auditing performance. 
    For instance, we can elect a leader that is most favorable for the adversary. 

    \item It simulates a client who submits requests periodically and a configurable number of nodes who replicate and commit log entries in response. 

    \item It is also capable of assigning the adversary to a node and simulating the fork and bad vote attacks in Examples \ref{eg:log-manip} and \ref{eg:obs-voting}. 
    It can launch the attack once at any time during request submissions.

    \item The simulator ensures the adversary is able to generate legitimate data to prevent it from being caught before consistency checks. 

    \item In the simulator, the nodes write their committed log entries in order into chunked files. 
    They also write their commitment certificate and leader certificates. 
\end{itemize}

\begin{figure*}[!htb]
    \centering
    \begin{subfigure}[b]{\linewidth}
        \includegraphics[width=\linewidth]{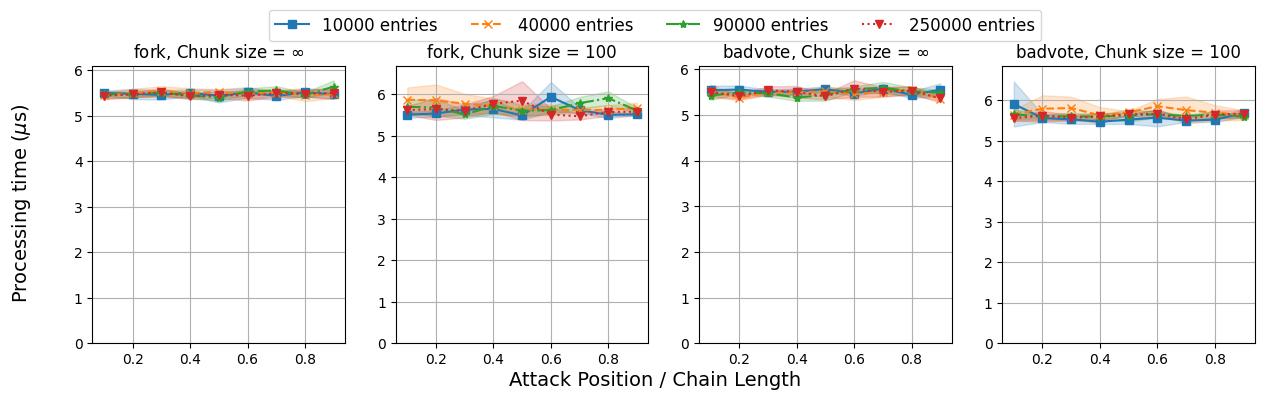}
        \caption{Processing time of \FIntegrate, the first stage of auditing.}
        \label{fig:offline-legit}
    \end{subfigure}
    \begin{subfigure}[b]{\linewidth}
        \includegraphics[width=\linewidth]{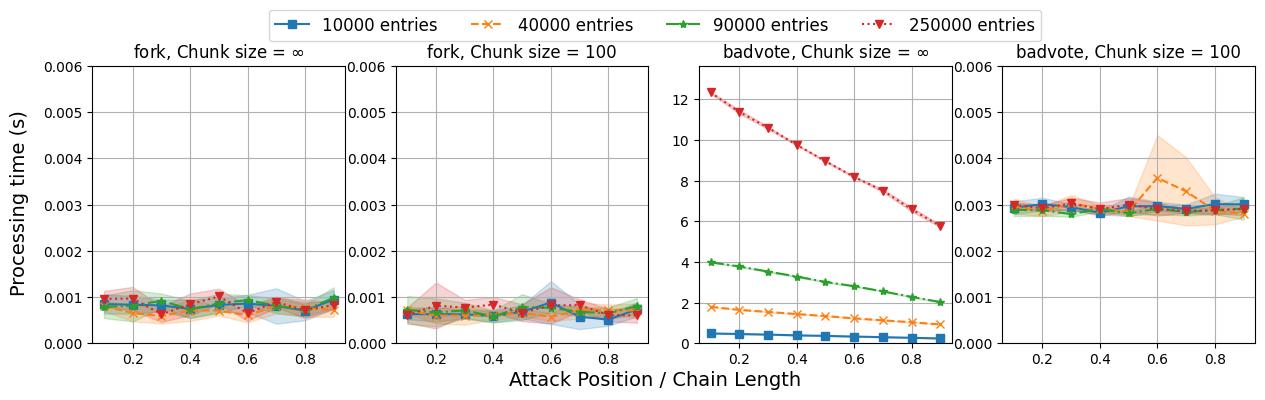}
        \caption{Processing time of \FAuditAll, the second stage of auditing.}
        \label{fig:offline-consistency}
    \end{subfigure}
    \caption{Evaluation of different stages of the auditing algorithm. }
\end{figure*}

Using this simulator, we set up five server nodes where one of them is Byzantine.
We generate transaction traffics of sizes 10,000, 40,000, 90,000 and 250,000. 
This results in approximately the same number of entries on the log chain of each node. 
We let the Byzantine node launch two types of attacks -- fork (Example \ref{eg:log-manip}) and bad vote (Example \ref{eg:obs-voting}). 
For each attack, we set two different chunk sizes $\infty$ and $100$, which corresponds to the max number of log entries in each file. 
Under each configuration of these parameters, we let the Byzantine node launch the attack at various positions during the transaction traffic. 
Specifically, the position can be expressed by the proportion of the time of attack to the time of the last transaction, which is selected from $0.1$ to $0.9$. 
We plot the time consumption of \FIntegrate in \fref{fig:offline-legit} and the time consumption of 
\FAuditAll in \fref{fig:offline-consistency}. 

By \fref{fig:offline-legit}, the time consumption of \FIntegrate is constant over each log entry, which implies a linear time complexity over all entries. 
This constant is irrelevant of the attack position or the size of the log list. 
Notably, the time consumption is improved when we choose chunk size $100$, instead of saving everything in the same file. 

By \fref{fig:offline-consistency}, the time consumption of \FAuditAll is constantly low when the Byzantine node launches the fork attack or the nodes use chunked files for storage. 
In the fork attack, it is easy for the auditor to discover from the last entries that the log lists conflict at the same term, and then expose the leader of that term without further looking into the depths of the lists. 
In the bad vote attack, however, the auditor must search for log entries by a given term. 
Specifically, when two nodes $u$ and $v$ have different ending terms $t_u$ and $t_v$ where $t_u > t_v$ without loss of generality, the auditor must find the first entry in $u$'s log list that has a higher term than $t_v$. 
The earlier the attack, the deeper this entry is, and the longer it takes for the auditor to scan from rear to front as we do in this experiment. 
This explains the third subfigure of \fref{fig:offline-consistency}.
When the log entries are stored in chunks, the auditor can derive the file that stores this entry and directly search within the file which has no more than $100$ entries. 
This technique accelerates consistency checks by up to $4000\times$ by comparing the third and fourth subfigures. 

With the total processing time reduced to milliseconds, the \FAuditAll algorithm for consistency checks spends much shorter time than the \FIntegrate algorithm for data legitimacy checks, which takes tens or hundreds of milliseconds. 
This is consistent with our complexity analysis in \sref{sec:eval-offline}.

\subsection{Dashboard of Auditing Algorithm}

\begin{figure*}
    \centering
\includegraphics[width=.9\linewidth]{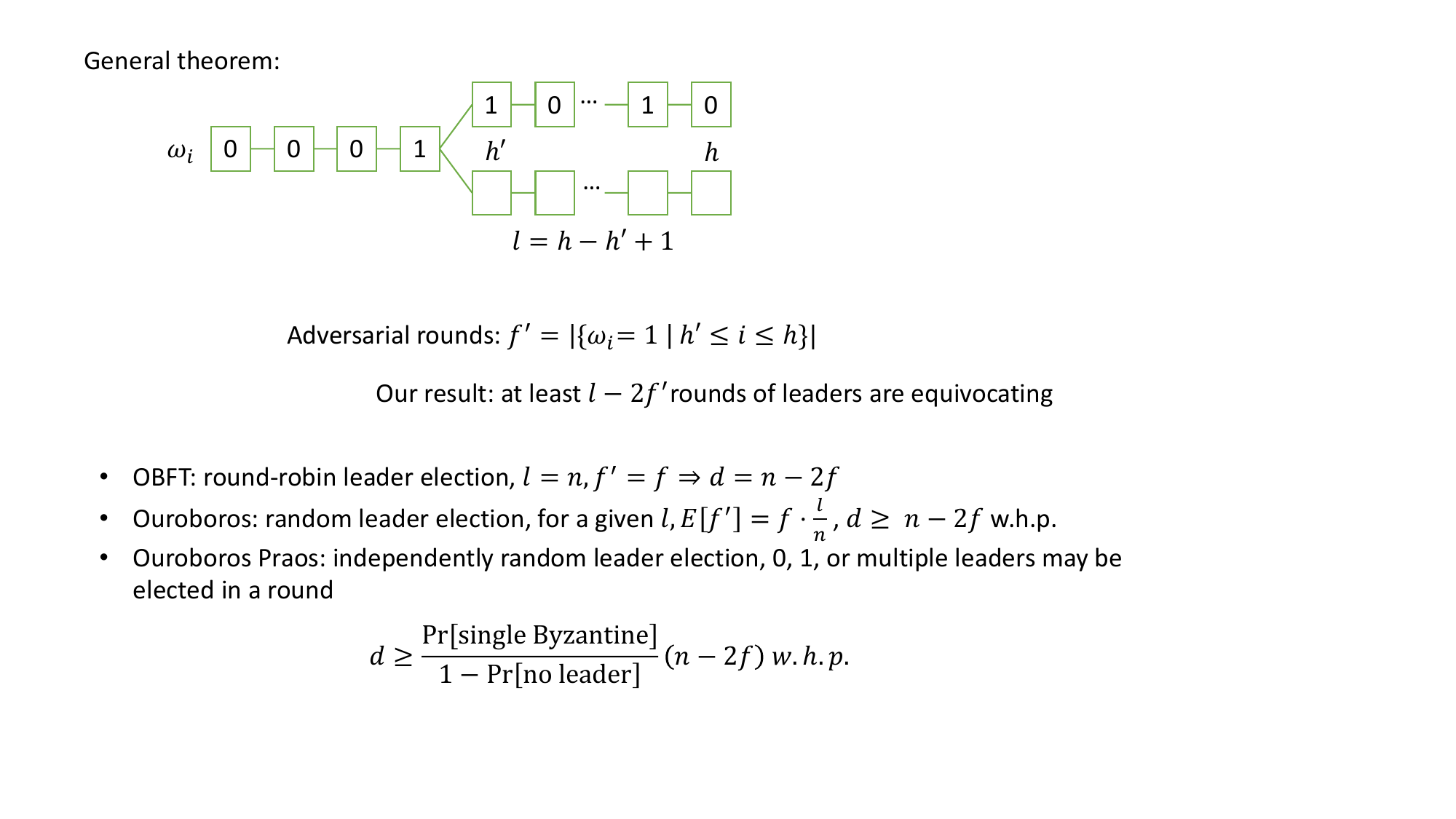}
    \caption{Dashboard of auditing in \sysname. This figure demonstrates a simulation of a blockchain system, consisting of five nodes with leader election occurring every 20 transactions. One of the nodes, node 4, is a Byzantine node that launches a bad vote attack at 70\% progress of the simulation, resulting in a safety violation. Upon detecting the conflict, the auditor process issues an alert and initiates an auditing algorithm to identify the culprit and extract evidence. In this particular scenario, node 4 is successfully detected, and the extracted evidence shows that it voted for a CC in term 3 and later voted for a conflicting LC in term 4. }
    \label{fig:demo}
\end{figure*}

\fref{fig:demo} shows the screenshot of the dashboard mentioned in \sref{sec:eval-offline}. 


\subsection{Comparison Against a BFT Protocol}
\label{sec:discussion-bft}

\begin{figure*}
    \centering
    \includegraphics[width=\linewidth]{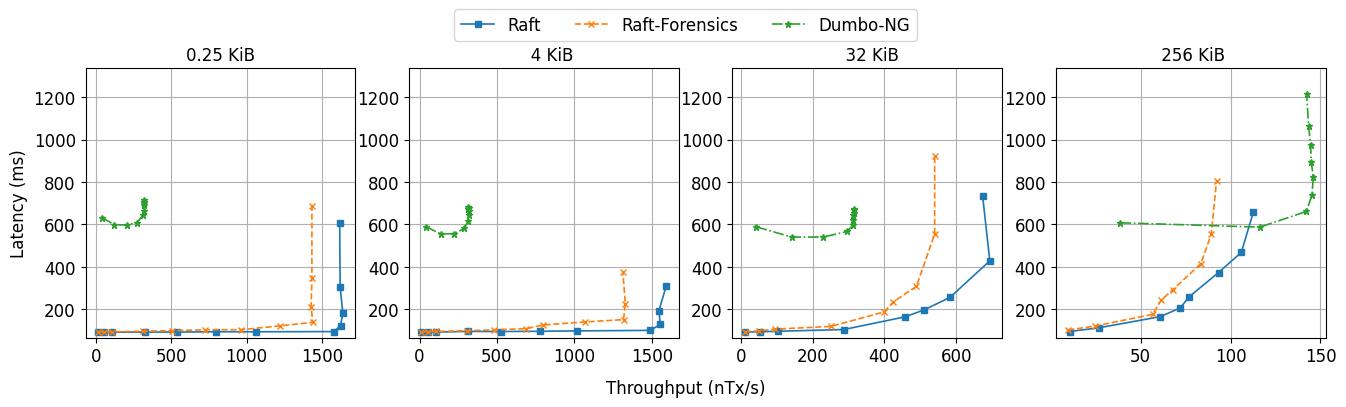}
    \includegraphics[width=\linewidth]{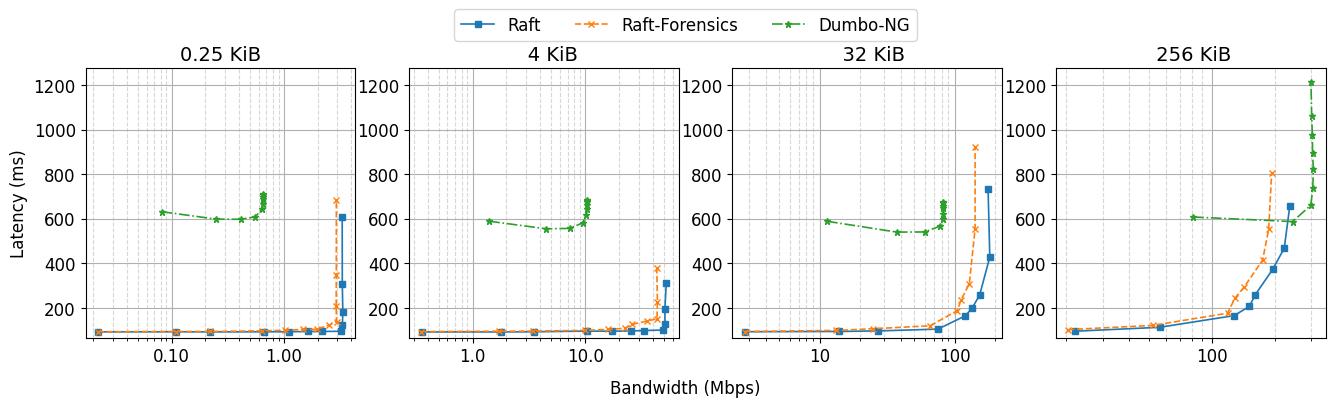}
    \caption{Latency-throughput tradeoff between CFT and BFT protocols ($n = 4$ nodes). Top row displays throughput in number of transactions per second; bottom row displays throughput in bandwidth.}
    \label{fig:bft}
\end{figure*}

Figure \ref{fig:bft} compares the performances of Raft-Forensics, Raft and Dumbo-NG. 
We chose the official implementation \cite{gitdumbong} for Dumbo-NG, where all servers spawn transactions by default. 
For a fair comparison between both Raft variants and Dumbo-NG, we disabled transaction batching for Dumbo-NG by setting batch sizes to 1 as batching is not implemented for Raft. 
It may have a negative impact on Dumbo-NG's throughput in number of transactions at small transaction sizes, but the throughput in bandwidth suffers less because high transaction sizes can simulate batching. 

This may not be the fairest comparison between the performances of CFT and BFT protocols -- it is ideal to compare Dumbo-NG in its default working state against Raft, which additionally incorporates batching and separated transaction dissemination. 
However, implementing these changes in Raft  would entail significant additional engineering in Raft, which conflicts with our objective of minimizing modifications. 
This objective is motivated by the widespread adoption of Raft and the associated costs and challenges of replacing or modifying it.

Under each configuration {of transaction size and client concurrency}, we run all the nodes and client processes simultaneously for 1 minute for Dumbo-NG which requires up to 40 seconds to warm up (20 seconds for Raft variants that do not require warm-up).
For the remaining configurations, we use the same settings as in \S\ref{sec:eval}.

\end{document}